\documentclass[conference]{IEEEtran}
\usepackage[cmex10]{amsmath}\interdisplaylinepenalty=2500
\usepackage{amssymb}
\usepackage{relsize} %for large \Rho math symbol
\usepackage{graphicx}
\usepackage{xspace}
\usepackage{stmaryrd}
\DeclareMathSymbol{\widehatsym}{\mathord}{largesymbols}{"62}
\DeclareMathSymbol{\widetildesym}{\mathord}{largesymbols}{"65}

\newcommand{\arXiv}[1]{\myurl{http://arXiv.org/abs/#1}}
\newcommand{\myurl}[1]{{\rm\texttt{#1}}\xspace}

\newcommand{\COMMENTED}[1]{}

\newcommand{\IR}{\mathbb{R}}
\newcommand{\ID}{\mathbb{D}}

\newcommand{\IP}{\mathbb{P}}

\newcommand{\IZ}{\mathbb{Z}}

\newcommand{\IN}{\mathbb{N}}
\newcommand{\dom}{\operatorname{dom}}

\newcommand{\Lip}{\operatorname{Lip}}
\newcommand{\Card}{\operatorname{Card}}

\newcommand{\graph}{\operatorname{graph}}

\newcommand{\id}{\operatorname{id}}

\newcommand{\sdzero}{\textup{\texttt{0}}\xspace}
\newcommand{\sdone}{\textup{\texttt{1}}\xspace}

\newcommand{\Oracle}{\mathfrak{O}}

\newcommand{\calA}{\mathcal{A}}

\newcommand{\calC}{\mathcal{C}}
\newcommand{\calF}{\mathcal{F}}
\newcommand{\calO}{\mathcal{O}}

\newcommand{\calK}{\mathcal{K}}
\newcommand{\calKC}{\mathcal{KC}}
\newcommand{\calKR}{\mathcal{KR}}

\newcommand{\calS}{\mathcal{S}}

\newcommand{\mycite}[2]{{\rm\cite[\textsc{#1}]{#2}}}

\newcommand{\mapstoto}{\Mapsto}
\newcommand{\diam}{\operatorname{diam}}
\newcommand{\Entropy}{\mathcal{E}}

\newcommand{\Round}{R}
\newcommand{\Sound}{S}

\newcommand{\ball}{B}
\newcommand{\cball}{\overline{\ball}}
\newcommand{\Sphere}{\calS}

\newcommand{\closure}[1]{\overline{#1}}

\newcommand{\myrho}{\varrho}

\newcommand{\Hom}{\operatorname{H}}
\newcommand{\Aut}{\operatorname{Aut}}
\newcommand{\Vol}{\operatorname{Vol}}
\DeclareMathOperator*{\Area}{Area}

\newcommand{\naive}{na\"{\i}ve\xspace}

\newcommand{\Hausdorff}[1]{{#1}_{\text{H}}}
\newcommand{\Supremum}[1]{{#1}_{_\infty}}

\def\undertilde#1{\mathord{\vtop{\ialign{##\crcr
$\hfil\displaystyle{#1}\hfil$\crcr\noalign{\kern1.5pt\nointerlineskip}
$\hfil\tilde{}\hfil$\crcr\noalign{\kern1.5pt}}}}}

\newtheorem{theorem}{Theorem}
\newtheorem{definition}[theorem]{Definition}

\newtheorem{fact}[theorem]{Fact}

\newtheorem{example}[theorem]{Example}

\newtheorem{remark}[theorem]{Remark}
\newtheorem{question}[theorem]{Question}

\begin{document}
\title{Computable Operations on Compact Subsets \\ of Metric Spaces
with Applications to \\ Fr\'echet Distance and Shape Optimization}

\author{\IEEEauthorblockN{Chansu Park\IEEEauthorrefmark{1}, 
Ji-Won Park\IEEEauthorrefmark{2}, 
Sewon Park\IEEEauthorrefmark{2}, 
Dongseong Seon\IEEEauthorrefmark{1} and 
Martin Ziegler\IEEEauthorrefmark{2}}
\IEEEauthorblockA{\IEEEauthorrefmark{1}Seoul National University, \qquad
\IEEEauthorrefmark{2}KAIST School of Computing \qquad (Republic of Korea)}}


\maketitle
\begin{abstract}
\boldmath
We extend the Theory of Computation on real numbers, continuous real functions, 
and bounded closed Euclidean subsets,
to compact metric spaces $(X,d)$: thereby generically including computational 
and optimization problems over higher types, such as the compact `hyper' spaces 
of (i) nonempty closed subsets of $X$ w.r.t. Hausdorff metric, 
and of (ii) equicontinuous functions on $X$. %wrt. supremum norm.
The thus obtained Cartesian closure is shown to exhibit the same structural
properties as in the Euclidean case, particularly regarding function pre/image.
This allows us to assert the computability of (iii) Fr\'{e}chet Distances 
between curves and between loops, as well as of (iv) constrained/Shape Optimization.
\end{abstract}
\IEEEpeerreviewmaketitle
\addtocounter{footnote}{2}

%%%%%%%%%%%%%%%%%%%%%%%%%%%%%%%%%%%%%%%%%%%%%%%%%%%%%%%%%%%%%%%%%%%%%%%%%%
\section{Introduction, Motivation, Background}
{Identifying} (and justifying) the `right' concepts and 
notions is crucial for the foundation of a theory.
The classical Theory of Computing is based on Turing machines
with data  encoded in binary and runtime taken in the worst-case 
over all inputs of length $n\to\infty$ asymptotically
--- for discrete data. The Theory of Computing with
continuous data also dates back to Turing (1937) 
for single reals and to Grzegorczyk (1957) for real functions;
yet the quest for the right notions
over higher types is still in progress \cite{Royer,SchroederHierarchy,AkiSTOC,Longley}
since an input here contains infinite information and 
cannot even be read in full before having to start producing output.
The present work continues the pursuit \cite{metric} for a uniform treatment
of computability and complexity on general compact metric spaces $(X,d)$:
thus generically including operations on the `higher type' spaces 
of (i) nonempty closed subsets of $X$ w.r.t. the Hausdorff distance,
and of (ii) equicontinuous functions from $X$ to another compact metric space $Y$
w.r.t. the supremum norm.
We are guided by the structural properties exhibited in Computational Logic \cite{Kreitz,Escardo}
and by the well-established Euclidean case \cite{Ko91,Brattka,Weihrauch}:

%%%%%%%%%%%%%%%%%%%%%%%%%%%%%%%%
\subsection{Computing Real Numbers, Functions, Closed Subsets}
\emph{Computing} a real number $r$ means to produce an %\emph{in}finite 
integer sequence $a_m$ of numerators of dyadic rationals $a_m/2^m$ 
approximating $r$ up to absolute error $\leq2^{-m}$.
And \emph{computing} a 
(possibly partial) function $f:\subseteq\IR^d\to\IR$ 
%with respect to the maximum norm 
means:
\begin{gather} 
%\begin{gathered}
\text{Convert any sequence $(\vec a_m)\subseteq\IZ^d$ satisfying} \label{e:Function} \\ 
\text{$|\vec x-\vec a_m/2^m|\leq2^{-m}$,
\quad $\vec x:=\lim\nolimits_m \vec a_m/2^m\in\dom(f)$,} \nonumber \\
\text{to some $(b_n)\subseteq\IZ$
s.t. $|y-b_n/2^n|\leq2^{-n}$ for $y=f(\vec x)$}  \nonumber
%\end{gathered}
\end{gather}
while the behaviour on other sequences $(\vec a_m)$ is arbitrary 
\cite[\S4.3]{Weihrauch}.
For example addition $\IR^2\to\IR$ is computed by converting 
$(\vec a_m)\in\IZ^2$ to $b_n:=\lfloor a_{n+2,x}/4+a_{n+2,y}/4\rceil$.
%and is obviously feasible in time polynomial in the output precision $n$.
A computation according to Equation~(\ref{e:Function}) runs \emph{in time $T(n)$}
if $b_n$ appears after at most $T(n)$ steps, regardless of $\vec x\in\dom(f)$ or $(\vec a_m)$.
Following \mycite{Definition~4.8}{Brattka} and \mycite{Exercise~5.2.1}{Weihrauch},
call a non-empty compact $W\subseteq\IR^d$ \emph{computable}
iff there exists a family $A_m\subseteq\IZ^d$ 
such that $W$ has Hausdorff distance $\leq2^{-m}$ 
to $\{\vec a_m/2^m : \vec a_m\in A_m\big\}$,
where $(A_m)$ is required to be \emph{uniformly recursive}
in the sense that 
$\prod_m \{m\}\times\bar A_m\subseteq \IN\times\IZ^{d}$
is decidable. $W$ is \emph{co-computable} 
if there exists a uniformly co-r.e. such family. We report:

\begin{fact} \label{f:Euclidean}
Fix non-empty compact $W\subseteq[0;1]^d$ 
and computable total $\Lambda:W\to\IR^e$.
\begin{enumerate}[\IEEEsetlabelwidth{}]
\item[a)]
$[0;1]^d$ itself is computable.
The union of two co-/ computable subsets is again co-/computable,
and the intersection of two co-computable sets is co-computable;
cmp. \mycite{Theorem~5.1.13}{Weihrauch}.
\item[b)]
A point $\vec x\in[0;1]^d$ is computable ~iff~
the compact singleton $\{\vec x\}\subseteq [0;1]^d$ is co-computable
iff $\{\vec x\}$ is computable;
cmp. \mycite{Example~5.1.12.1}{Weihrauch}.
\item[c)]
If $W$ is computable, then it contains some computable point;
cmp. \mycite{Exercise~5.1.13}{Weihrauch}.
\item[d)] 
$\Lambda$ admits a runtime bound $T=T(n)$, i.e., depending
only on the output precision $n$; %but not on the argument $\vec x\in W$
%nor on the approximating sequence $(\vec a_n)\subseteq\IZ^d$;
and for any such bound $T$, $n\mapsto T(n+1)+1$ is a \emph{binary modulus of continuity} 
in that $|\vec x-\vec{x}'|\leq2^{-T(n+1)}$ implies
$\big|\Lambda(\vec x)-\Lambda(\vec{x}')\big|\leq2^{-n+1}$;
cmp. \mycite{Theorem~2.19}{Ko91} and \mycite{Theorem~7.2.7}{Weihrauch}.
\item[e)]
If $W$ is co-computable, then the set \quad
$\displaystyle \big\{ \big(\vec a_0,\ldots,\vec a_n \big) \;:$
\[  n\in\IN, \; \exists \vec w\in W \;
   \forall j\leq n: \; \vec a_j\in\IZ^d \;\wedge\; |\vec a_j/2^j-\vec w|\leq 2^{-j} \big\} \]
(of finite initial sequences of dyadic sequences converging to some $\vec w\in W$) is co-r.e.
and $\Lambda$ has a recursive %binary modulus of continuity and 
runtime bound; cmp. \mycite{Theorems~2.4.7+7.2.5+7.2.7}{Weihrauch}.
\item[f)]
If $W$ and non-empty compact $V\subseteq[0;1]^e$ are co-computable, 
then so is $\Lambda^{-1}[V]\subseteq W$;
cmp. \mycite{Example~5.1.19.2}{Weihrauch}.
\item[g)]
If $W$ is computable, then
the image $\Lambda[W]\subseteq[0;1]^e$ is again computable compact
\mycite{Example~5.2.11}{Weihrauch}.
\item[h)]
If compact $W\subseteq[0;1]^d$ coincides with the closure of its interior, $\overline{W^\circ}$,
and both $W$ and $[0;1]^d\setminus R^\circ$ are co-computable,
then they are computable %{\rm\cite{MLQ1}}.
\mycite{Theorem~3.1}{MLQ2}.
\item[j)]
If non-empty compact $W\subseteq[0;1]$ is computable,
then so are $\max W\in[0;1]$ and $\min W\in[0;1]$;
cmp. \mycite{Lemma~5.2.6}{Weihrauch}.
%\item[x)] If $\Lambda:W\to\IR^e$ is injective, 
%then its inverse $Y\ni\vec y\mapsto f^{-1}(\vec y)\in X$ is computable
%\mycite{Theorem~6.3.11}{Weihrauch}
\end{enumerate}
\end{fact}
Item~d) follows from careful continuity and compactness considerations.
It corresponds to, and generalizes the (trivial) observation 
in discrete complexity theory
that any total computation on $\{\sdzero,\sdone\}^*$ 
admits a worst-case runtime bound depending only on the
length $n$ of, but not on the input $\vec x\in\{\sdzero,\sdone\}^n$ itself.
This complexity-theoretic property in turn is the key to prove Items~e) to h)
although the latter are only concerned with computability.

Item~a) is optimal in that there exist computable compact $V,W\subseteq[0;1]$ 
such that $V\cap W$ is not computable \mycite{Exercise~5.2.11}{Weihrauch};
and, regarding Item~f), there exists a computable $\Lambda:[0;1]\to[0;1]$
such that $\Lambda^{-1}[0]\neq\emptyset$ contains no computable point
\mycite{Exercise~6.3.12}{Weihrauch}.
Also Ernst Specker constructed a recursive and
increasing sequence of integer fractions whose supremum
is not computable \mycite{Example~1.3.2}{Weihrauch}.

The former condition is known as lower semi-computability
\mycite{Definition~2}{Buchin} or left-computability;
in fact a real number $x$ is computable iff it is 
both left and right computable \mycite{Lemma~4.2.5}{Weihrauch}. 
We also record that, with the notation from Equation~(\ref{e:Function}), 
strict inequality ``$f(x)>0$'' is equivalent to
``$\exists n: b_n>1$'' and thus r.e. (recursively enumerable, aka semi-decidable),
but in general undecidable \mycite{Exercise~4.2.9}{Weihrauch}.
We use \emph{computable} for the continuous realm,
\emph{decidable}/\emph{co-/recursive/enumerable}
for the discrete one.

%%%%%%%%%%%%%%%%%%%%%%%%%%%%%%%%%%%%%%%%%%%
\subsection{Overview, Previous and Related Work}
The present work generalizes the Theory of Computation
from Euclidean unit cubes to compact metric spaces $(X,d)$.
Being separable, computing here naturally means approximation up to error $2^{-n}$
by a sequence (of indices w.r.t. a fixed partial enumeration $\xi:\subseteq\IN\to X$)
of some countable dense subset,
thus generalizing the dyadic rationals $\ID=\{a/2^n:a,n\in\IZ\}$ 
canonically employed the real case;
cmp. \cite[\S2]{PER89} or \mycite{Definition~8.1.2}{Weihrauch}.
Of course the particular choice of said enumeration $\xi$ heavily affects 
the computational properties it induces \cite{Presser,SchroederComplexity}.

We propose in Definition~\ref{d:Metric} weak conditions on $\xi$
that assert the entire Fact~\ref{f:Euclidean} to carry over;
see Theorem~\ref{t:Metric}. They permit the categorical
construction of dense enumerations, again satisfying said conditions,
for (i) the compact space of non-empty closed subsets of $X$ 
equipped with the Hausdorff distance,
and for (ii) the compact space of equicontinuous functions 
from $X$ to another compact metric space,
identified with their graph: Theorem~\ref{t:Exp}.
We demonstrate the relevance and applicability of these conditions
by asserting (Theorem~\ref{t:Frechet})
the computability of the Fr\'{e}chet Distance
between curves and between loops;
and by asserting 
computability of the generic nonlinear optimization problem
$\max\{ \Lambda(x) : \Phi(x)\leq 0 \}$
for every computable cost function $\Lambda:X\to\IR$
and every computable, feasible, and open constraint $\Phi:X\to\IR$
including the contemporary case of Shape Optimization: Theorem~\ref{t:Opt}.

\medskip
%\paragraph{Previous and Related Work}
The so-called \emph{Type-2 Theory of Effectivity} %standard reference on Computable Analysis
considers computability on second countable topological $\operatorname{T}_0$ spaces
\mycite{Lemma~3.2.6}{Weihrauch} by means of partial encodings as infinite binary 
sequences, that is, over Cantor space. It establishes Cartesian closure by
constructing generic encodings of countable products \mycite{Definition~3.3.3}{Weihrauch} 
and of the space of (relatively) continuous functions \mycite{Definition~3.3.13}{Weihrauch}
as well as, for the case of a complete metric space, of its induced Hausdorff hyperspace
\mycite{Exercise~8.1.10}{Weihrauch}. 
However, lacking (local/sigma) compactness, properties (d)+(f)+(h) 
from Fact~\ref{f:Euclidean} do not hold in general.
Indeed several notions of computability, equivalent in the Euclidean case \cite{Brattka},
have been shown distinct for separable metric spaces \cite{Presser}.
In fact, compactness %, in its various constructive variants \cite{Oliva}
is well-known crucial in Pure as well as in Computable Analysis \cite{SchroederComplexity,Escardo,Steinberg}.

%%%%%%%%%%%%%%%%%%%%%%%%%%%%%%%%%%%%%%%%%%%%%%%%%%%%%%%%%%%%%%%%%%%%%%%%%%%%%%%%%%%%%%%
\subsection{Recap: Continuous Functions and Compact Metric Spaces} \label{ss:Math}

We presume a basic comprehension of mathematical calculus 
and properties of compact metric spaces $(X,d)$.
Write $\ball(x,r)=\{ x'\in X : d(x,x')<r\}$ 
for the open ball in $X$ with center $x\in X$ and radius $r\geq0$,
$\cball(x,r)$ for its closure (unless $r=0$).
More generally abbreviate $\ball(S,r):=\bigcup_{x\in S}\cball(x,r)$
and similarly for $\cball(S,r)$, $S\subseteq X$.
$S^\circ$ denotes the interior (=largest open subset), 
$\bar S$ the closure (=least closed superset) of $S$;
$\partial S:=\closure{S}\setminus S^\circ$ is the boundary of $S$.
Write $\cball^d:=\big\{\vec x\in\IR^d : x_1^2+\cdots+x_d^2\leq1\big\}$
for the closed Euclidean $d$-dimensional unit ball,
$\Sphere^{d-1}:=\partial\cball^d$ for the unit sphere.
Let $\calC(X,Y)$ denote the space of continuous functions $f:X\to Y$;
$\calC(X)$ in case $Y=\IR$. %The image of $f:\subseteq X\to Y$ 
%on $W\subseteq X$ is written $f[W]=\{f(w):w\in W\cap\dom(f)\}\subseteq Y$; 
%pre-image $f^{-1}[V]\subseteq X$ for $V\subseteq Y$.

\begin{definition} \label{d:Math}
\begin{enumerate}[\IEEEsetlabelwidth{}]
\item[a)]
A \emph{modulus of continuity} of $f:X\to Y$ 
is a non-decreasing right-continuous 
mapping $\omega:[0;\infty)\to[0;\infty)$
such that $\omega(t)\to0$ as $t\to0$ and
$e\big(f(x),f(x')\big)\leq\omega\big(d(x,x')\big)$
holds for all $x,x'\in X$.
\item[b)]
A \emph{binary modulus of continuity} of $f:X\to Y$
is a non-decreasing mapping $\mu:\IN\to\IN=\{0,1,\ldots\}$ such that
$e\big(f(x),f(x')\big)\leq2^{-n}$ holds whenever $d(x,x')\leq2^{-\mu(n)}$. 
\item[c)]
Abbreviate
\quad $
\calC_\mu(X,Y) \;=\; \big\{ f:X\to Y \::\: f \text{ has binary modulus of continuity } \mu \big\}
$; $\Lip_1:=\calC_{\id}$ is the space of non-expansive functions,
$\Lip_2:=\calC_{n\mapsto n+1}$.
%We omit $Y$ in case $Y=\IR$.
\item[d)]
Let $\IN^\IN=\{ (v_0,v_1,\ldots) : v_n\in\IN \}$ denote Baire space,
equipped with the metric 
$\beta(\bar v,\bar w)=2^{-\min\{n:v_n\neq w_n\}}$.
\item[e)]
For $\emptyset\subsetneq V,W\subseteq X$ consider 
the distance function $d_V:X\ni x\mapsto\inf\{ d(x,v) : v\in V\}\geq0$
and Hausdorff distance
$\Hausdorff{d}(V,W)=\max\big\{\sup\{ d_V(w) :w\in W\},\sup\{d_W(v):v\in V\}\big\}$.
The sup.metric on $\calC(X,Y)$ is denoted
$\Supremum{e}(f,g)=\max\big\{ e\big(f(x),g(x)\big): x\in X\big\}$.
\item[f)]
$\Entropy_X:\IN\to\IN$ denotes Kolmogorov's metric \emph{entropy},
also known as \emph{modulus of uniform boundedness} \mycite{Def~18.52}{Kohlenbach}:
It is defined such that $X$ can be covered by $2^{\Entropy_X(n)}$
open balls of radius $2^{-n}$, but not by $2^{\Entropy_X(n)-1}$.
\end{enumerate}
\end{definition}
If the entropy grows linearly, its asymptotic slope coincides 
with the \emph{Minkowski-Bouligand} or \emph{box-counting dimension} of $X$;
otherwise the latter is infinite. Compare also \cite{metric,Mayordomo}\ldots
The central mathematical tool of the present work is compactness,
so let us recall some aspects of this concept relevant in the sequel:

\begin{fact} \label{f:Compact}
Suppose $(X,d)$ is a compact metric space;
i.e., every sequence in $X$ admits a convergent subsequence; 
equivalently: every cover 
$\bigcup_n \ball(x_n,r_n)\supseteq X$ by open balls contains
a finite subcover $\bigcup_{n\leq N} \ball(x_n,r_n)\supseteq X$.
Fix another compact metric space $(Y,e)$.
\begin{enumerate}[\IEEEsetlabelwidth{}]
\item[a)]
$X$ is complete, that is, every Cauchy sequence converges.
A non-empty subset of $X$ is closed ~iff~ the restriction of $d$ turns it into a compact space.
Every continuous $f:X\to\IR$ attains its infimum and supremum.
\item[b)]
Every $f\in\calC(X,Y)$ has compact image $:=f[X]\subseteq Y$,
is uniformly continuous and thus admits a (binary) modulus of continuity.
\item[c)]
By the \emph{Arzel\`a Ascoli Theorem}, 
a set $\calF\subseteq\calC(X,Y)$ has compact closure
~iff~ it is a subset of $\calC_\mu(X,Y)$
for some binary modulus of continuity $\mu$.
\item[d)]
By \emph{K\"onig's Lemma}, a non-empty subset $C\subseteq\IN^\IN$ of Baire space
has compact closure ~iff~ each node has finite degree
in the tree of finite initial segments
\[ C^* :=\; \big\{ \vec u=(u_0,\ldots u_{n}) \:\big|\: n\in\IN, \:
\exists \bar v: \: \vec u\circ\bar v\in C \big\} \;\subseteq\; \IN^* \]
\item[e)]
The set $\calK(X)$ of non-empty closed subsets of $X$ 
equipped with the Hausdorff distance $\Hausdorff{d}$ from Definition~\ref{d:Math}f)
constitutes again a compact metric space. %If $X$ is convex, then so is $\calK(X)$.
%Let $\calK_0(X):=\calK(X)\cup\{\emptyset\}$ with $\emptyset$ as isolated point.
We shall call $\big(\calK(X),\Hausdorff{d}\big)$ \emph{the} Hausdorff hyper-space over $X$.
\item[f)] 
Fix $W\in\calK(X)$,
$D:\IN\to\IN$ a sequence, and $\xi:\subseteq\IN\to X$ some (possibly partial) enumeration. 
Then the subsets $x_{\xi,D}=\prod_m x_{\xi,D,m}$ 
and $W_{\xi,D}:=\bigcup_{x\in W} x_{\xi,D}$ 
of Baire space $\IN^\IN$ are compact, where we use
the abbreviations
$x_{\xi,D,m}:=\big\{ u\in [2^{D(m)}]\cap\dom(\xi) : d\big(x,\xi(u)\big)\leq2^{-m}\big\}$
and $[M]:=\{0,1,\ldots,M-1\}$.
\item[g)]
Consider the compact metric space $(X\times Y,d\times e)$, where 
$(d\times e)\big((x,y),(x',y')\big):=\max\big\{d(x,x'),e(y,y')\big\}$. 
A total function $f:X\to Y$ is continuous iff
$\graph(f):=\big\{\big(x,f(x)\big):x\in X\big\}$
is compact, i.e. an element of $\calK(X\times Y)$.
\item[h)]
Moreover it holds \;$\displaystyle
\Hausdorff{(d\times e)}\big(\graph(f),\graph(g)\big) \leq$
\[ \;\leq\; \Supremum{e}(f,g) \;\leq\; 
(\omega+\id)\Big(\Hausdorff{(d\times e)}\big(\graph(f),\graph(g)\big)\Big)
\]
for $\omega$ any modulus of continuity of $f$ or $g$;
and $\calF\subseteq\calC(X,Y)$ is compact iff
$\graph(\calF):=\{\graph(f):f\in\calF\}\subseteq\calK(X\times Y)$ is.
\end{enumerate}
\end{fact}
Item~a) justifies the minimum and maximum in Definition~\ref{d:Math}f).

%%%%%%%%%%%%%%%%%%%%%%%%%%%%%%%%%%%%%%%%%%%%%%%%%%%%%%%%%%%%%%%%%%%%%%%%
\section{Computing on a Compact Metric Space} \label{s:Metric} %\label{ss:Compact}
Here we extend Fact~\ref{f:Euclidean}
from Euclidean $[0;1]^d$ to arbitrary compact metric spaces.
Generalizing the real case with dyadic rationals $\ID$
as canonical countable dense subset, computation on a metric space is commonly defined 
by operating on (sequences of indices wrt.) some fixed countable partial dense enumeration;
cmp. \cite[\S2]{PER89} or \mycite{Definition~8.1.2}{Weihrauch}.
Of course, the particular choice of said enumeration heavily affects 
whether, and which items of, Fact~\ref{f:Euclidean} carry over \cite{SchroederComplexity}.
In Definition~\ref{d:Metric} below we propose conditions that 
formalize and generalize numerical (i.e. Euclidean) grids 
to (i) more general compact metric spaces
by considering a relaxation of \emph{Hierarchical Space Partitioning}
whose (ii) subsets are closed balls that however 
(iii) may overlap as long as 
(iv) their centers keep distance $\eta$ from each other:

\begin{definition} \label{d:Metric}
Fix a compact metric space $(X,d)$ of diameter
$\diam(X):=\max\{d(x,x'):x,x'\in X\}\leq 1$.
\begin{enumerate}[\IEEEsetlabelwidth{}]
\item[a)]
For $m\in\IZ$, an \emph{$m$-covering of $X$} is a subset $X_m\subseteq X$ 
such that $X\supseteq\bigcup_{x\in X_m} \cball(x,2^{-m-1})$.
For $\eta\in\IN$, $X_m\subseteq X$ is \emph{$\eta$-separated} if it holds
$d(x,x')\geq2^{-\eta}$ for all distinct $x,x'\!\in\! X_m$;
%$X_m$ is 
\emph{$\eta$-rectangular} if $\forall x,x'\!\in\! X_m:2^\eta\cdot d(x,x')\!\in\!\IN$.% for all $x,x'\in X_m$.
\item[b)]
$(X,d,\xi,D)$ is a \emph{presented (compact) metric space} 
if $\xi:\subseteq\IN\to X$ is a partial dense enumeration 
and $D:\IN\to\IN$ strictly increasing such that,
for every $m\in\IN$, the image $\xi\big[[2^{D(m)}]\big]$ of $[2^{D(m)}]\cap\dom(\xi)$
constitutes an $m$-covering of $X$. 
%\\
For strictly increasing $\eta:\IN\to\IN$ and injective $\xi$,
$(X,d,\xi,D)$ is \emph{$\eta$-rectangular/$\eta$-separated}
if, for every $m\in\IN$, $\xi\big[[2^{D(m)}]\big]$ 
constitutes an $\eta(m)$-rectangular/$\eta(m)$-separated $m$-covering of $X$, respectively.
\item[c)]
Presented metric space $(X,d,\xi,D)$ is \emph{computably compact} 
if $\dom(\xi)$ and $D:\IN\to\IN$ are recursive and the following is semi-decidable:
\qquad $\displaystyle \big\{ (a,b,n,u,v) \;\big|\; u,v\in\dom(\xi),$
\[ a/2^n < d\big(\xi(u),\xi(v)\big) < b/2^n \big\}\subseteq\IN^5 
\]
%$(X,d,\xi,D)$ is \emph{computably $\eta$-rectangular} 
%if in addition the following mapping is well-defined and recursive:
%\begin{equation} \label{e:Rectangular} 
% \IN^3 \;\supseteq\; \bigcup\nolimits_m \big\{m\big\}\times 
%  \big([2^{D(m)}]\cap\dom(\xi)\big)^2  % \times  \big([2^{D(m)}]\cap\dom(\xi)\big)
%  \;\ni\; (m,u,v) \;\mapsto\; 2^{\eta(m)}\cdot d\big(\xi(u),\xi(v)\big) \;\in\; \IN 
%\end{equation}
%If the mappings
%$\IN\ni 2^m\mapsto\ell(m)$ and $\IN\ni2^m\mapsto \eta(m)$ are
%computable in time polynomial in the binary/bit-length of the input
%(for $2^m$ equivalent to: in $m$) and if so is the mapping from Equation~(\ref{e:Rectangular}) 
%(which requires its domain to be polynomial-time decidable and $\eta$ to grow at most polynomially),
%then the $\eta$-rectangular presented space $(X,d,\xi,D)$ is \emph{polynomial-time computable};
%similarly for exponential time, for logarithmic, and for polynomial space. XXX
\item[d)] 
For presented $(X,d,\xi,D)$, a \emph{name} of a point $x\in X$
is a sequence $\bar u=(u_m)$, $u_m\in\dom(\xi)\cap[2^{D(m)}]$,
with $d\big(\xi(u_m),x\big)\leq2^{-m}$. 
The point $x$ is \emph{computable} if it admits a recursive name.
It is \emph{polynomial-time computable}
if there exists a Turing machine which prints a name $\bar u$
such that $u_m$ appears after a number of steps 
bounded by some polynomial in $m$.
\item[e)]
Fix presented $(X,d,\xi,D)$ and $(Y,e,\upsilon,E)$
and recall that $W_{\xi,D}^*\subseteq\dom(\xi)^*$ 
denotes the set of finite initial sequences of elements $\bar u\in W_{\xi,D}$.
A \emph{name} of a partial mapping $\Lambda:\subseteq X\to Y$ 
is a (w.r.t. initial substrings) monotonic, total mapping 
$\Lambda^*:X_{\xi,D}^*\to Y_{\upsilon,E}^*$
such that, for every $\bar u\in x_{\xi,D}$ with $x\in\dom(\Lambda)$,
the (w.r.t. initial substrings) non-decreasing sequence $\big(\Lambda^*(u_0,\ldots,u_m)\big)_m$ 
has unbounded length and supremum $\bar v\in\Lambda(x)_{\upsilon,E}$.
We write $\Lambda^*_n(\bar u)$ for the $n$-th element $v_n$ of said supremum.
%\\
\emph{Computing} $\Lambda$ means 
(for a Turing machine with write-only right-moving tape) 
to compute some name $\Lambda^*$.
Such a computation runs in time $t(n)$ %and space $s(n)$ 
if it takes at most that many steps to output $\Lambda^*_n(\bar u)$,
independently of $\bar u\in\dom(\Lambda)_{\xi,D}$. 
\item[f)] 
For presented $(X,d,\xi,D)$, a \emph{name}
of compact non-empty $W\subseteq X$ is a sequence $\bar A=(A_m)$ of finite sets 
$A_m\subseteq[2^{D(m)}]\cap\dom(\xi)$ such that, 
for every $m\in\IN$, the set $\xi[A_m]\subseteq X$
has Hausdorff distance (Definition~\ref{d:Math}f) at most $2^{-m}$ to $W$.
The empty sequence is a name of $\emptyset$.
%\\
$W$ is \emph{computable} if it has a name $\bar A=(A_m)$
which is uniformly recursive in the sense that the set
$\prod_m \{m\}\times A_m\subseteq\IN\times\IN$ is decidable.
If said set is co-r.e., $W$ is called \emph{co-computable}.
%$W$ is computable in \emph{polynomial time} if
%there exists a sequence $u_m\in A_m$ computable in time polynomial in $m$
%and if $\{(2^m,u):u\in A_m\}\subseteq\IN^2$ is decidable in time 
%polynomial in the input length, that is, in $m+1+\lb(u)$
%where $\lb(N):=\lceil\log_2(1+N)\rceil$.
%\\
A \emph{standard} name $\bar A$ of $W$ satisfies 
$d_W\big(\xi(a)\big)<2^{-m}$ for every $a\in A_m\subseteq[2^{D(m)}]\cap\dom(\xi)$,
and 
$d_W\big(\xi(a)\big)>2^{-m-1}$ for every $a\in[2^{D(m)}]\cap\dom(\xi)\setminus A_m$.
\item[g)]
A \emph{rounding function} for presented $(X,d,\xi,D)$ 
is a mapping $\Round:\dom(\xi)\times\IN\to\dom(\xi)$ such that it holds
$\Round(u,m)\in[2^{D(m)}]$ and $d\Big(\xi\big(\Round(u,m)\big),\xi(u)\Big)\leq2^{-m-1}$.
%We measure its computational cost in terms of $m+1+\lb(u)$, 
%i.e. considering $m$ given in unary.
%For $u\in\dom(\xi)$ abbreviate $\bRound(u):=\big(\Round(u,m)\big)_{_m}\in\xi(u)_{\xi,D}$.
%\item[h)] XXX 2nd order representation?
\end{enumerate}
\end{definition}
\mycite{Definition~6.2.2}{WeihrauchComplexity} calls an $m$-covering \emph{$(m+1)$-spanning}.
Various common notions of computability for closed subsets, 
equivalent over $\IR^d$ \mycite{Theorem~3.6}{Presser},
%\mycite{Lemmas~5.1.7+5.1.10}{Weihrauch} 
are known to become distinct over more general spaces 
%already in terms of computability
\mycite{Theorems~3.9(3)+3.11(4)+3.15(2)}{Presser}.
Definition~\ref{d:Metric} thus has been crafted with great care.
For instance, although computations according to e) 
operate on approximations up to absolute error $\leq2^{-m}$, 
the requirement in a) of an $m$-covering
to provide strictly better approximations is crucial.
Item~c) strengthens \mycite{Definition~8.1.2.3}{Weihrauch}
in requiring $\dom(\xi)$ to be recursive, thus asserting
a computably compact space $X$ to be a computable subset of itself
in the sense of Item~f); see Theorem~\ref{t:Metric}a) below.
It also guarantees the set $X_{\xi,D}^*\subseteq\IN^*$ to be co-r.e.;
see Theorem~\ref{t:Metric}e).
%It asserts the restricted distance function 
%\begin{equation} \label{e:Distance}
%\IN\times\IN \;\supseteq\; \dom(\xi)\times\dom(\xi) \;\ni\; (u,v) \;\mapsto\; d\big(\xi(u),\xi(v)\big) \;\in\; \IR
%\end{equation}
%to be computable in the sense of Definition~\ref{d:Metric}e)
The rest of this subsection will provide further justification
by comparison to the Euclidean Fact~\ref{f:Euclidean}.
Indeed, Definition~\ref{d:Metric} generalizes the real case:

\begin{example} \label{x:Spaces}
\begin{enumerate}[\IEEEsetlabelwidth{}]
\item[a)]
Let $\ID_m:=\{a/2^m: a\in\IN, 0\leq a<2^m\}$ and $\ID:=\bigcup_m\ID_m$
denote the set of dyadic rationals in $[0;1)$.
Define $D(m):=m+1$ as well as $\myrho(0):=0$ and inductively
$\myrho:[2^{m+1}]\setminus[2^m]\ni a+2^{m}\mapsto (2a+1)/2^{m+1}\in\ID_{m+1}\setminus\ID_{m}$. 
%Extend this total enumeration to dimension $d\in\IN$ with $D^{(d)}:=d\cdot D$
%and $\myrho^{(d)}:\big\{2^{md},\ldots,2^{(m+1)d}-1\big\}\to\ID^d_{m+1}\setminus\ID^d_m$.
%Then $\big([0;1]^d,\;|\:\cdot\:|,\;\myrho^{(d)},d\cdot(\id+1)\big)$ 
Then $\big([0;1],\;|\:\cdot\:|,\;\myrho,\id+1\big)$ 
constitutes a computably $m$-rectangular (formally: $\id$-rectangular)
%logarithmic-space/polynomial-time computably 
compact space. It admits a %logarithmic-space/polynomial-time 
computable rounding function, namely 
\begin{multline*}
\Round(\cdot,m) \;: \; [2^{m+n+1}]\setminus[2^{m+n}] \;\ni\; a+2^{m+n} \;\mapsto\\
\mapsto\;
2^m + \lfloor (2a+1)/(2^{n+1})-\tfrac{1}{2}\rceil \;\in\; [2^{m+1}] \enspace .
\end{multline*}
\item[b)]
The `circle' $[0;1)\mod 1$ with the metric $d(x,y)=\min\{|x-y|,|x-y-1|\}$,
equipped with the enumeration $\myrho$ from (a) but now taking $D(m)=m$, 
is also computably $\id$-rectangular compact.
\item[c)]
Consider Cantor space $\{\sdzero,\sdone\}^\IN$ equipped
with the metric $\beta$ inherited from Baire space; 
recall Definition~\ref{d:Metric}d).
We turn this into a computably $m$-rectangular 
%logarithmic-space/polynomial-time computably 
compact space 
$\big(\{\sdzero,\sdone\}^\IN,\beta,\gamma,\id+1\big)$ as follows:
Let $\tilde\gamma:\IN\to\{\sdzero,\sdone\}^*$ 
enumerate all finite binary strings in order of length, 
define $\gamma(0):=\sdzero^\omega$, 
$\gamma(u+1):=\tilde\gamma(u)\circ\sdone\circ\sdzero^\omega$;
and truncation constitutes a %logarithmic-space/polynomial-time 
computable rounding function.
\item[d)]
Suppose $\psi:X\to Y$ is a homeomorphism 
between computably compact $(X,d,\xi,D)$ 
and topological space $Y$.
Then $\big(Y,d\circ\psi^{-1},\psi\circ\xi,D\big)$
constitutes a computable compact metric space
(thus justifying the notation in b).
If the former is $\eta(m)$-separated/rectangular/has
rounding function $R$, then so does the latter.
\item[e)]
For computably compact metric spaces $(X,d,\xi,D)$ and $(Y,e,\upsilon,E)$,
their Cartesian product
$(X\times Y,d\times e,\xi\times\upsilon,D\cdot E)$ becomes again a
computably compact metric space
by defining $\xi\times\upsilon:\subseteq\IN\to X\times Y$ inductively 
on $\big[D(m)\cdot E(m);D(m+1)\cdot E(m+1)-1\big]\cap\IN$: 
\begin{align*}
%\begin{split}
D(m)\cdot E(m)+\big(D(m+1)-D(m)\big)\cdot v+u  &\;\mapsto\\[-0.2ex]
\mapsto\; \Big(\xi\big(D(m)+u\big),\upsilon\big(E(&m)+v\big)\Big), \\
\IN\ni u<D(m+1)-D(m), \quad \IN\ni v&<E(m)
\end{align*}\begin{align*}% \\[0.6ex]
D(m+1)\cdot E(m)+D(m+1)\cdot v+u &\;\mapsto \\[-0.2ex]
\mapsto \; \Big(\xi\big(u\big),\upsilon\big(E(&m)+v\big)\Big), \\
\IN\ni u<D(m+1), \quad \IN\ni v<E(m+1)&-E(m) 
\end{align*}
If $X$ and $Y$ are $\eta(m)$-separated/rectangular,
then so is $X\times Y$. Recursive rounding functions
for $X$ and $Y$ give rise to one for $X\times Y$.
%Extend this total enumeration to dimension $d\in\IN$ with $D^{(d)}:=d\cdot D$
%and $\myrho^{(d)}:\big\{2^{md},\ldots,2^{(m+1)d}-1\big\}\to\ID^d_{m+1}\setminus\ID^d_m$.
%Then $\big([0;1]^d,\;|\:\cdot\:|,\;\myrho^{(d)},d\cdot(\id+1)\big)$ 
\end{enumerate}
\end{example}
Items~a+e) generalize the case of real vectors.
A computable compact subset need not in turn
constitute a computably compact metric space:
consider for example $\{1/\pi\}\subseteq[0;1]$.

\begin{remark} \label{r:Spaces}
\begin{enumerate}[\IEEEsetlabelwidth{}]
\item[a)] 
To every compact $(X,d)$ with partial dense enumeration $\xi:\subseteq\IN\to X$,
there exists some $D:\IN\to\IN$ 
rendering $(X,d,\xi,D)$ a presented metric space. 
If the restricted distance 
\[ %\begin{equation} \label{e:Distance}
\IN\times\IN \supseteq \dom(\xi)\times\dom(\xi) \ni (u,v) \mapsto d\big(\xi(u),\xi(v)\big) \in \IR
\] %\end{equation} according to Equation~\ref{e:Distance} 
is computable and the natively $\Pi_2$ set
\begin{align}
\Big\{ &\big(u,n,v_1,n_1,\ldots v_j,n_j\big) \;\big|\; j\in\IN, \; u,v_1,\ldots v_j\in\dom(\xi), 
\nonumber \\
  &\cball\big(\xi(u),2^{-n}\big) \subseteq \ball\big(\xi(v_1),2^{-n_1}\big) \cup\cdots
     \ball\big(\xi(v_j),2^{-n_j}\big) \Big \} \nonumber \\
=& \Big\{ \big(u,n,v_1,\ldots n_j\big) \:\big|\:
  \forall w\in\dom(\xi): d\big(\xi(w),\xi(u)\big)\!>\!2^{-n}
\nonumber \\ \label{e:Covering}
& \qquad \qquad \vee\; \exists i\leq j: 
\; d\big(\xi(w),\xi(v_i)\big)<2^{-n_i} \: \Big\}  %\;\subseteq\; \IN^*
\end{align}
is actually semi-decidable, then there exists a recursive such $D$.
\mycite{Definition~2.6}{Presser} calls Equation~(\ref{e:Covering}),
with dyadic radii generalized to arbitrary rationals,
the \emph{effective covering property}. 
\item[b)]
Every compact metric space $(X,d)$ admits a partial dense enumeration $\xi$
such that $D(m):=\Entropy_X(m+1)$ turns $(X,d,\xi,D)$ into 
an $(m+1)$-separated space: Let $\xi[2^{D(m)}]$ enumerate the
centers of open balls of radius $2^{-m-1}$ covering $X$.
\\
Conversely whenever $(X,d,\xi,D)$ is $\eta$-separated, 
it holds $D(m)\leq\Entropy_X\big(\eta(m)+1\big)$:
Consider some choice of $2^{\Entropy_X(\eta(m)+1)}$ centers
of open balls of radius $2^{-\eta(m)-1}$ covering $X$;
then each one can contain at most one of the points 
in $\xi[2^{D(m)}]$, since the latter have pairwise 
distance $\geq2^{-\eta(m)}$; cmp. \cite[\S3]{metric}.
\item[c)]
Not every convex compact metric space admits a rectangular enumeration, though:
Consider the geodesic distance on a circle of irrational circumference.
Even in the Euclidean case,
it has been conjectured since Erd\"os and Ulam (1946) that no open subset 
of $\IR^2$ admits a dense sequence of pairwise rational distances;
cmp. \myurl{http://terrytao.wordpress.com/2014/12/20}
\item[d)]
Suppose $(X,d,\xi,D)$ is computably compact. Then computably compact
$\big(X,d,\xi,m\mapsto D(m+1)\big)$ admits a recursive rounding function:
Given $u\in\dom(\xi)$ and $m\in\IN$ there exists, and 
Definition~\ref{d:Metric}c) asserts
enumeration of $[2^{D(m+1)}]\cap\dom(\xi)$ to find,
some $v=:\Round(u,m)$ with $d\big(\xi(v),\xi(u)\big)<2^{-m-1}$. 
(In $[2^{D(m)}]\cap\dom(\xi)$, distance $2^{-m-1}$ is feasible, 
but not necessarily computably so\ldots)
\item[e)]
$(A_m)$ is a name of non-empty compact $W\subseteq X$ iff (i)
every $\bar a=(a_m)$ with $a_m\in A_m$ satisfying
\begin{equation} \label{e:Name}
 \forall n,m: d\big(\xi(a_m),\xi(a_n)\big)\leq2^{-m}+2^{-n}
\end{equation}
constitutes a name of some $x\in W$ and if (ii) conversely to every $x\in W$ there exists
a name $\bar a$ such that Equation~(\ref{e:Name}) holds: 
If $(A_m)$ constitutes a name of $W$,
then (i) every $\bar a$ satisfying Equation~(\ref{e:Name})
gives rise to $x:\lim_m\xi(a_m)\in X$ by completeness,
and $d_W\big(\xi(a_m)\big)\leq2^{-m}$ shows $x\in W$;
and (ii) to $x\in W$ and every $m\in\IN$ there exists
some $a_m\in A_m$ with $d\big(\xi(a_m),x\big)\leq2^{-m}$,
hence satisfying Equation~(\ref{e:Name}).
%Conversely if every $\bar a$ satisfying Equation~(\ref{e:Name})
%constitutes a name of some $x\in W$, then $A_m\subseteq\dom(\xi)\cap[2^{D(m)}]$
%and $d_W\big(\xi(a_m)\big)\leq2^{-m}$; while the existence of such $\bar a$
%to every $x\in W$ implies $d\big(\xi(a_m),x\big)\leq2^{-m}$.
\item[f)]
Every standard name of some non-empty compact $W\subseteq X$ is also a name of $W$: 
By Definition~\ref{d:Metric}b) there exists,
to every $x\in W$, some $a\in\dom(\xi)\cap[2^{D(m)}]$
with $d\big(\xi(a),x\big)\leq2^{-m-1}$;
and $(A_m)$ being a standard name of $W$ requires
$a\in A_m$ whenever $d_W\big(\xi(a)\big)\leq2^{-m-1}$:
thus $(A_m)$ is a name of $W$. 
\\
Conversely, to every co-r.e/recursive name $(A_m)$ of non-empty 
compact $W\subseteq X$, there exists a co-r.e./recursive 
\emph{standard} name of $W$: Indeed every $A'_m$ with
\begin{multline} \label{e:standard}
\Big\{ a'\in\dom(\xi)\cap[2^{D(m)}] \;\big|\;
  \exists a\in\dom(\xi)\cap[2^{D(m+3)}]:  \\
\quad a\in A_{m+3} \;\wedge\; d\big(\xi(a),\xi(a')\big)\pmb{\leq5}\cdot2^{-m-3} \Big\}
\;\subseteq \\ \subseteq \; A'_m \;\subseteq
\; \Big\{ a'\in\dom(\xi)\cap[2^{D(m)}] \;\big|\;
  \exists a\in A_{m+3} \;\wedge\qquad \\
\;\wedge\; d\big(\xi(a),\xi(a')\big)\pmb{<7}\cdot2^{-m-3} \Big\} 
\end{multline}
constitutes such a standard name:
To $a\in A_{m+3}$ there exists some $x\in W$ with $d\big(\xi(a),x\big)\leq2^{-m-3}$,
hence $a'\in A'_m$ implies $d\big(\xi(a),\xi(a')\big)<7\cdot2^{-m-3}$ 
and in turn $d_W\big(\xi(a')\big)<(7+1)\cdot2^{-m-3}=2^{-m}$; 
while $a'\in \dom(xi)\cap[2^{D(m)}]\setminus A'_m$ implies
$d\big(\xi(a),\xi(a')\big)>5\cdot2^{-m-3}$ for every $a\in A_{m+3}$,
and in turn $d_W\big(\xi(a')\big)>(5-1)\cdot 2^{-m-3}=2^{-m-1}$.
Now recall that strict inequality of distances is r.e.,
and non-strict is co-r.e.
Hence, if $(A_m)$ is uniformly co-r.e., 
then so is the left-hand side of Equation~(\ref{e:standard});
and if $(A_m)$ is even uniformly recursive,
then the right-hand side is uniformly r.e.:
now apply the next item to its complement.
\item[g)]
Fix pairwise disjoint families $X_m$ and $Z_m$
of uniformly co-r.e. subsets of integers.
Then there exists a uniformly recursive family
$Y_m$ disjoint to $Z_m$ with $X_m\subseteq Y_m$: \\
For given $x,m$ search in parallel for a witness
that $x\not\in X_m$ and for one that $x\not\in Z_m$
and report the (negation of the) first to succeed.
%since $(\IN\setminus X_m)\cup(\IN\setminus Z_m)=\IN$.
\item[h)]
Suppose $(X,d,\xi,D)$ and $(Y,e,\upsilon,E)$ are computably compact
with rounding function $\Round:\dom(\xi)\times\IN\to\dom(\xi)$.
Let $(A_m)_{_m}$ denote a name of non-empty compact $W\subseteq X$, 
and $\Lambda^*$ a name of $\Lambda:W\subseteq X \to Y$,
computable in time $t(n)$. Recall that $W_{\xi,D}^*\subseteq\IN^*$ 
denotes the set of all finite initial segments of sequences in $W_{\xi,D}$
and observe that, for every $n\in\IN$ and $a\in A_{t(n)}$,
even though $\xi(a)$ itself might not even lie in $W$, 
the closed ball $\cball\big(\xi(a),2^{-t(n)}\big)$ 
does intersect $W$. Moreover, for every $x\in W\cap\cball\big(\xi(a),2^{-t(n)}\big)$, 
the finite sequence
$\vec a:=\big(\Round(a,0),\Round(a,1),\ldots,\Round(a,t(n)-1),a\big)$
extends to some name $\bar a$ of $x$, i.e., belongs to $W_{\xi,D}^*$.
Abbreviating $y:=\Lambda(x)$, the machine computing $\Lambda^*$ will
on input $\bar a$ produce $\Lambda^*_n(\bar a)=:b_n\in[\dom(\upsilon)\cap[2^{E(n)}]$
with $e\big(\upsilon(b_n),y\big)\leq2^{-n}$ for $y:=\Lambda(x)$.
However, in time bound $t(n)$ it cannot even read past $\vec a$;
hence $\Lambda^*_n(\bar a)$ depends only on $\vec a$:
$\Lambda^*_n(\bar a)=\Lambda^*_n(\bar u)$ holds
whenever $\beta(\bar a,\bar u)\leq2^{-t(n)}$. \\
Let us denote by $\Lambda^t_{n}(a)$ the thus well-defined composite mapping 
$A_{t(n)}\ni a\mapsto\vec a\mapsto\bar a\mapsto\Lambda^*_n(\bar a)\in[\dom(\upsilon)\cap[2^{E(n)}]$
satisfying $e\Big(\upsilon\big(\Lambda^t_n(a)\big),y\Big)\leq2^{-n}$
for all $y\in\Lambda\big[W\cap\cball\big(\xi(a),2^{-t(n)}\big)\big]$.
\IEEEQED\end{enumerate}\end{remark}
Item~g) can be regarded as 
a discrete counterpart to Fact~\ref{f:Euclidean}j).
Item~h) is based on the continuity/adversary argument
underlying the sometimes so-called \emph{Main Theorem} 
%of Computable Analysis 
\cite[\S2.2]{Weihrauch},
Fact~\ref{f:Euclidean}d+e), and Theorem~\ref{t:Metric}d+e) below.

%%%%%%%%%%%%%%%%%%%%%%%%%%%%%%%%%%%%%%%%%%%%%%%%%%%%%%
\subsection{Computable Operations on a Compact Metric Space} 
For presented metric spaces,
the composition of two computable functions is
again computable; computable functions map computable points
to computable points; a constant function is computable
iff its value is computable. Moreover,
similarly to Fact~\ref{f:Euclidean} and \mycite{Def~4.1}{WeihrauchComplexity}, we have:

\begin{theorem} \label{t:Metric}
Let $(X,d,\xi,D)$ and $(Y,e,\upsilon,E)$ be computably compact spaces
with recursive rounding functions $\Round$ and $\Sound$
according to Definition~\ref{d:Metric}g).
Suppose compact non-empty $W\subseteq X$ and total $\Lambda:W\to Y$ are computable.
\begin{enumerate}[\IEEEsetlabelwidth{}]
\item[a)]
$X$ is a computable subset of itself.
The union of two co-/computable sets is again co-/computable;
the intersection of two co-computable sets is co-computable.
\item[b)] 
A point $x\in X$ is computable ~iff~
the compact singleton $\{x\}\subseteq X$ is co-computable
iff $\{x\}\subseteq X$ is computable.
\item[c)]
If $W$ is computable, it contains some computable point.
\item[d)] 
$\Lambda$ admits a computable time bound $T=T(n)$ depending
only on the output precision $n$; 
for any such bound $T$,
$n\mapsto T(n\!+\!1)\!+\!1$ 
is a binary modulus of continuity 
of $\Lambda$.
\item[e)]
If $W$ is co-computable, then $W_{\xi,D}^*$ is co-r.e.; 
and $\Lambda$ has a recursive binary modulus of continuity and runtime bound.
\item[f)] 
If $W$ and non-empty compact $V\subseteq Y$ are co-computable, then so is $\Lambda^{-1}[V]\subseteq W$.
\item[g)]
%Suppose $X$ and $Y$ admit recursive rounding functions according to Definition~\ref{d:Metric}g).
If non-empty compact $W$ is computable, then 
the image $\Lambda[W]\subseteq Y$ is again (compact and) computable.
\item[h)]
If non-empty compact $W\subseteq X$ coincides with $\overline{R^\circ}$
and both $W$ and $X\setminus R^\circ$ are co-computable,
then they are computable.
\item[j)]
A total $\Lambda:W\to Y$ is computable (in the sense of Def.\ref{d:Metric}e)
~iff~ the compact set $\graph(\Lambda)\subseteq X\times Y$ is computable. 
\end{enumerate}
\end{theorem}
Item~g) asserts, together with Fact~\ref{f:Euclidean}j),
that $\max\Lambda[W]$ and $\min\Lambda[W]$ are computable reals 
for every computable non-empty compact $W\subseteq X$ and $\Lambda:W\to[0;1]$.
And Items~b+f) imply that, under the same hypothesis
and for computable $y\in Y$,
a \emph{unique} solution $x\in W$ to the equation ``$\Lambda(x)=y$''
is computable. Item~j) effectivizes Fact~\ref{f:Compact}g),
that is the identification of a function as a transformation
with its graph as a `static' object, and 
justifies our encoding of compact function spaces
in Subsection~\ref{ss:Exp} below; cmp \cite[\S4]{Braverman}.
%\pagebreak

%%%%%%%%%%%%%%%%%%%%%%%%%%%%%%%%%%%%%%%%%%%%%%%
\subsection{Proof of Theorem~\ref{t:Metric}} 
\begin{enumerate}[\IEEEsetlabelwidth{}]
\item[a)]
By Definition~\ref{d:Metric}b),
$A_m:=\dom(\xi)\cap[2^{D(m)}]$ is a name of $X$;
and uniformly recursive according to Definition~\ref{d:Metric}c).
\\
For $A_m,B_m\subseteq[2^{D(m)}]\cap\dom(\xi)$ with 
$\Hausdorff{d}\big(\xi[A_m],W\big),\Hausdorff{d}\big(\xi[B_m],V\big)\leq2^{-m}$,
$A_m\cup B_m\subseteq[2^{D(m)}]\cap\dom(\xi)$ has
$\Hausdorff{d}\big(\xi[A_m\cup B_m],W\cup V\big)\leq2^{-m}$;
and is uniformly recursive/co-r.e. whenever both $(A_m)$ and $(B_m)$ are.
The cases $W=\emptyset$ or $V=\emptyset$ are easily treated separately.
Finally, for uniformly co-r.e. $A_n,B_{n'}$ subsets of recursive $\dom(\xi)\cap[2^{D(m)}]$,
\begin{multline*} C_m \;:=\; \big\{ c\in A_m \;\big|\; \forall n,n'\in\IN \;
\exists a\in A_n \;\exists b\in B_{n'}: \\
  d\big(\xi(a),\xi(b)\big)\leq2^{-n}+2^{-n'}
  \wedge\;   d\big(\xi(c),\xi(a)\big)\leq 2^{-n}+2^{-m} \big\}
\end{multline*}
is co-r.e. (since $\dom(\xi)$ is) and a name of $V\cap W$:
To every $x\in V\cap W$ there exist $a_n\in A_n$ and $b_{n'}\in B_{n'}$
with $d\big(\xi(a_n),x\big)\leq2^{-n}$ and $d\big(\xi(b_{n'}),x\big)\leq2^{-n'}$,
so $d\big(\xi(a_n),\xi(b_{n'})\big)\leq2^{-n}+2^{-n'}$; 
and, since $x\in V$, there exists some $c\in A_m$ with $d\big(\xi(c),x\big)\leq2^{-m}$,
so $d\big(\xi(c),\xi(a)\big)\leq 2^{-n}+2^{-m}$: 
resulting in $c\in C_m$.
Conversely, to every $c\in C_m$, there are sequences 
$a_n\in A_n$ and $b_{n'}\in B_{n'}$, and thus $v_n\in V$
and $w_{n'}\in W$ with 
$d\big(\xi(a_n),v_n\big),d\big(\xi(b_{n'}),w_{n'}\big)\leq2^{-n'}$;
by compactness $\xi(a_{n_k})\to\lim_k v_{n_k}=:v\in V$ 
and $\xi(b_{n'_k})\to\lim_k w_{n'_k}=:w\in W$ 
for some subsequences; 
now $d\big(\xi(a_{n_k}),\xi(b_{n'_k})\big)\leq2^{-n_k}+2^{-n'_k}$
implies $v=w=:x\in V\cap W$; finally
$d\big(\xi(c),x\big)\leftarrow d\big(\xi(c),\xi(a_{n_k})\big)\leq 2^{-n_k}+2^{-m}\to 2^{-m}$
as $k\to\infty$.
\item[b)]
For computable $x$ with recursive name $\bar u=(u_m)_{_m}\in x_{\xi,D}$, 
the uniformly recursive sequence of singletons $A_m:=\{u_m\}$ constitutes
a name of $\{x\}$.
Conversely suppose $(A_m)$ is a co-r.e. name of $\{x\}$.
Then the sets 
\begin{multline*}
A'_m \;:=\;
\big\{ \Round(a',m) \;\big|\; a'\in\dom(\xi)\cap[2^{D(m+3)}], \\ 
%\; \forall a\in[2^{D(m+3)}]: \\ 
\forall a\in[2^{D(m+3)}]: \; a\not\in A_{m+3} \;\vee\; d\big(\xi(a),\xi(a')\big)<2^{-m-2} \big\}
\end{multline*}
are (i) uniformly semi-decidable, (ii) non-empty, and (iii)
any sequence $a'_m\in A'_m$ constitutes a name of $x$.
Indeed (ii) to $x\in X$ there exists $a'\in\dom(\xi)\cap[2^{D(m+3)}]$
with $d\big(\xi(a'),x\big)\leq2^{-m-4}$, while every 
$a\in A_{m+3}$ satisfies $d\big(\xi(a),x\big)\leq2^{-m-3}$.
Conversely (iii) every $a\in A_{m+3}$ satisfies $d\big(\xi(a),x\big)\leq2^{-m-3}$;
hence any $\Round(a',m)\in A'_m$ has $d\big(\xi(a'),x\big)<3\cdot2^{-m-3}$
and $d\big(\xi(\Round(a',m)),x\big)<7^{-m-3}$.
\item[c)]
Let $A_m\subseteq[2^{D(m)}]\cap\dom(\xi)$ be a uniformly recursive
name of compact non-empty $W\subseteq X$.
Then, starting with any $a_0\in A_2$,
one can iteratively computationally search for,
and according to Remark~\ref{r:Spaces}e) 
is guaranteed to find, some $a_{m}\in A_{m+2}$ with 
$d\big(\xi(a_{m}),\xi(a_n)\big)<2^{-m-1}+2^{-n-1}$ for all $n<m$:
recall that strict inequality of distances
is semi-decidable according to Definition~\ref{d:Metric}b).
Then $u_m:=\Round(a_m,m)\in\dom(\xi)\cap[2^{D(m)}]$ satisfies 
$d\big(\xi(u_{m}),\xi(u_n)\big)<2^{-m}+2^{-n}$ and
$d_W\big(\xi(u_m)\big)\leq 3\cdot2^{-m-2}<2^{-m}$
by triangle inequality; hence $x:=\lim_m\xi(u_m)\in W$.
\item[d)]
Fix $n\in\IN$ and recall from Remark~\ref{r:Spaces}h) that
a machine $\calA$ computing a name $\Lambda^*$ of $\Lambda$,
when presented with any sequence $\bar u=(u_m)_{_m}\in w_{\xi,\Xi}$ for some $w\in W$,
produces according to Definition~\ref{d:Metric}c) some
$v_n=\Lambda_n\in\dom(\upsilon)$ with $e\big(\Lambda(x),\upsilon(v_n)\big)\leq2^{-n}$:
after a finite number $T=T_{\calA}(n,\bar u)$ of steps and in particular `knowing' no more
than the first $T$ entries of $\bar u$.
The thus defined function $T(n,\cdot):W_{\xi,\ell}\to\IN$ (implicitly depending on $\calA$)
is therefore locally constant, that is, continuous:
$T(n,\bar u)=T(n,\bar u')$ whenever $\beta(\bar u,\bar u')\leq2^{-T(n,\bar u)}$.
\\
Now Fact~\ref{f:Compact}f) asserts $W_{\xi,\Xi}\subseteq\IN^\IN$
to be compact; hence, by Fact~\ref{f:Compact}a), $T(n,\cdot)$ is 
bounded by some least integer $T(n)$ (again depending also on $\calA$). 
We show that $n\mapsto T(n+1)+1$ constitutes a binary modulus of continuity of $\Lambda\big|_W$:
Fix $x\in W$ and consider for each $m\in\IN$ some $u_m\in[\Xi(m)]$ 
with $d\big(x,\xi(u_m)\big)\leq2^{-m\pmb{-1}}$
according to Definition~\ref{d:Metric}a).
For this particular $\bar u\in x_{\xi,\Xi}$,
every $x'\in\cball(x,2^{-m-1})$ has
$d\big(x',\xi(u_m)\big)\leq2^{-m}$ for all $m'\leq m$;
indeed $x'$ admits a sequence $\bar u'\in x'_{\xi,\Xi}$
which coincides with $\bar u$ on the first $m$ entries.
And for $m\geq T(n)\geq T(n,\bar u)$ by definition, $\calA$'s output $\bar v$
on input $\bar u$ coincides up to position $n$
with its output $\bar v'$ on input $\bar u'$.
Triangle inequality thus yields
\[ d(x,x')\leq2^{-T(n)-1} \quad\Rightarrow\quad
e\big(\Lambda(x),\Lambda(x')\big)\leq2^{-n+1} \enspace . \]
\item[e)]
Let $(A_m)_{_m}$ denote a co-r.e. name for $W$. 
According to Remark~\ref{r:Spaces}e), $W_{\xi,\ell}^*$ coincides with the set
\begin{multline} \label{e:Initial}
\Big\{ \vec u=(u_0,\ldots,u_{n}) \;\big|\; n\in\IN, \; \forall i,j\leq n:  \\
u_j\in\dom(\xi)\cap[2^{\ell(j)}] \:\wedge\;
d\big(\xi(u_i),\xi(u_j)\big)\leq2^{-i}+2^{-j} \wedge \\
\forall m \;\exists a_m\in A_m: \; d\big(\xi(a_m),\xi(u_j)\big)\leq2^{-m}+2^{-j} 
\Big\} 
\end{multline}
which is clearly co-r.e. with ``$\forall m$'' as only unbounded quantifier
and co-r.e. inequality ``$d\big(\xi(u_i),\xi(u_j)\big)\leq2^{-i}+2^{-j}$''.
Indeed, for every $\vec u=(u_0,\ldots,u_{n})$ according to Equation~(\ref{e:Initial})
and fixed $j\leq n$, there exists a sequence $a_m\in A_m$ such that,
one the one hand, $d_W\big(\xi(a_m)\big)\leq2^{-m}\to0$ hence
$\lim_n \xi(a_{m_n})=:x\in W$ for some subsequence;
while on the other hand
$2^{-j}\leftarrow 2^{-m_n+1}+2^{-j}\geq d\big(\xi(a_{m_n}),\xi(u_j)\big)\to d\big(x,\xi(u_j)\big)$.
This asserts $\vec u$ to extend to some $\bar u\in x_{\xi,\ell}$.
\\
To find a recursive time bound $T'\geq T$ to (d),
extend the partial algorithm $\calA$ computing $\Lambda^*$ from compact $W_{\xi,\Xi}$
to $\calA'_n$ accepting inputs $\bar u$ from the entire 
set $\prod_{m\geq 0}[2^{\ell(m)}]$ by, 
simultaneously to executing $\calA(\bar u)$ until it prints the $n$-th output symbol,
trying to refute $\bar u\in W_{\xi,\Xi}$ just established as co-r.e.
Noting that $\calA'_n$ indeed terminates on all possible 
inputs from compact $\prod_{m\geq 0}[2^{\ell(m)}]\subseteq\IN^\IN$
and therefore (d) in some time bound $T'(n)\geq T(n)$ depending only on $n$,
the following algorithm computes such $T'(n)$:
\begin{quote}
Initialize $T':=1$.
Simulate $\calA'_n$ on each $\vec u\in\prod_{m=0}^{T'-1}[2^{\ell(m)}]\subseteq\IN^{T'}$ until
either (i) it terminates or (ii) reads past the finite input.
In case (ii), increase $T'$ and restart; else output $T'$ and terminate.
\end{quote}
\item[f)]
First observe that, since $\Lambda$ is continuous according to (d), $\Lambda^{-1}$ maps
compact/closed $V\subseteq Y$ to a closed/compact subset of $W$ and/or $X$.
Now let $(A_m)_{_m}$ denote a co-r.e. name of $W$,
$(B_k)_{_k}$ similarly one of $V$,
and $\mu:\IN\to\IN$ simultaneously
a recursive both binary modulus of continuity of $\Lambda$ 
and runtime bound according to (e).
Now consider the uniformly
recursive mapping $\Lambda^\mu_n:A_{\mu(n)}\to\dom(\upsilon)\cap[2^{E(n)}]$
according to Remark~\ref{x:Spaces}h), and let
\begin{multline*} C_m \;:=\; \big\{ a\in\dom(\xi)\cap[2^{D(m)}] \;\big|\;
\forall n,n' \;\exists a'\in A_{\mu(n')} \\
\exists b\in B_n: \; 
e\big(\upsilon\big(\Lambda^\mu_{n'}(a')\big),\upsilon(b)\big)\leq2^{-n}+2^{-n'} \;\wedge \\
\wedge\; 
d\big(\xi(a),\xi(a')\big)\leq2^{-m}+2^{-\mu(n')} \big\} \enspace .
\end{multline*}
It is clearly uniformly co-r.e., since $(A_{\mu(n')})$ and $(B_n)$ are, and non-strict inequality is as well,
and both $\dom(\xi)\cap[2^{D(\mu(n'))}]$ and $\dom(\upsilon)\cap[2^{E(n)}]$ are recursive by hypothesis.
We show that $(C_m)$ constitutes a standard name of $\Lambda^{-1}[V]$:
Every $a\in\dom(\xi)\cap[2^{D(m)}]$ with $d_{\Lambda^{-1}[V]}\big(\xi(a)\big)\leq2^{-m-1}$
belongs to $C_m$: To $x\in\Lambda^{-1}[V]$ with $d\big(x,\xi(a)\big)]\leq2^{-m-1}$
and every $n,n'\in\IN$ there exist $a'\in A_{\mu(n')}$ with $d\big(x,\xi(a')\big)\leq2^{-\mu(n')}$
as well as $b\in B_n$ with $e\big(y,\upsilon(b)\big)\leq2^{-n}$ for $y:=\Lambda(x)\in V$;
hence $d\big(\xi(a),\xi(a')\big)\leq2^{-m-1}+2^{-\mu(n')}$,
and $e\big(\upsilon\big(\Lambda^\mu_{n'}(a')\big),y\big)\leq2^{-n'}$ 
implies $e\big(\upsilon\big(\Lambda^\mu_{n'}(a')\big),\upsilon(b)\big)\leq2^{-n'}+2^{-n}$.
\\
Conversely to $a\in C_m$ there exist sequences $a'_{n'}\in A_{\mu(n')}$
and $b_n\in B_n$ with 
$d\big(\xi(a),\xi(a'_{n'})\big)\leq2^{-m}+2^{-\mu(n')}$ and
$e\big(\upsilon\big(\Lambda^\mu_{n'}(a'_{n'})\big),\upsilon(b_n)\big)\leq2^{-n}+2^{-n'}$.
By compactness, $\lim_k\xi(a'_{n'_k})=x\in W$ 
and $\lim_k\upsilon(b_{n_k})=y\in V$ for some subsequences.
It follows $d\big(\xi(a),x\big)\leq2^{-m}$ as well as 
$e\big(\Lambda(x),y\big)\leftarrow
e\big(\upsilon\big(\Lambda^\mu_{n'_k}(a'_{n'_k})\big),\upsilon(b_{n_k})\big)\leq2^{-n_k}+2^{-n'_k}\to0$
as $k\to\infty$.
Therefore $x\in\Lambda^{-1}[y]$ and 
$d\big(\xi(a),x\big)\leftarrow d\big(\xi(a),\xi(a'_{n'_k})\big)\leq2^{-m}+2^{-\mu(n')}\to 2^{-m}$
as $k\to\infty$.
\item[g)] 
The image $\Lambda[W]\subseteq Y$ is compact by Fact~\ref{f:Compact}b).
To see its computability, 
fix a recursive name $(A_m)_{_m}$ of $W$,
a recursive binary modulus of continuity $\mu:\IN\to\IN$ of $\Lambda$ according to (e),
and uniformly recursive $\Lambda^\mu_{n}:A_{\mu(n)}\to\dom(\upsilon)\cap[2^{E(n)}]$
%with $e\Big(\upsilon\big(\Lambda^\mu_{n}(a')\big),\Lambda\big(\xi(a')\big)\big)\leq2^{-n}$
according to Remark~\ref{r:Spaces}h).
Now consider the set
\[ B_m \;:=\; \big\{ \Lambda^{\mu}_{m}(a) \;\big|\; a\in A_{\mu(m)} \big\}  \] 
which is clearly uniformly decidable since $\mu$ and $A_{\mu(m)}$ is.
We show that $(B_m)$ constitutes a name of $\Lambda[W]$:
To every $y=\Lambda(x)$ with $x\in W$ there exists by hypothesis
some $a\in A_{\mu(m)}$ with $d\big(\xi(a),x\big)\leq2^{-\mu(m)}$;
hence $e\Big(\upsilon\big(\Lambda^\mu_{m}(a)\big),y\Big)\leq2^{-m}$
by choice of $\mu$ and $\Lambda^{\mu}_{m}$.
Conversely, every $\Lambda^\mu_{m}(a)\in B_m$ 
arises from some $a\in A_{\mu(m)}$ 
and in turn some $x\in W$ with $d\big(\xi(a),x\big)\leq2^{-\mu(n)}$;
hence $e\Big(\upsilon\big(\Lambda^{\mu}_{m}(a)\big),y\Big)\leq2^{-m}$
for $y:=\Lambda(x)\in\Lambda[W]$.
\item[h)]
By Remark~\ref{r:Spaces}f) 
suppose w.l.o.g. that $(A_m)$ is a co-r.e. \emph{standard} name of $W$
and $(B_n)$ one of $X\setminus R^\circ$.  
We show that every family $(C_m)$ with 
\begin{multline} \label{e:Complement}
A_m \;\subseteq\; C_m \;\subseteq\;
\big\{ u\in\dom(\xi)\cap[2^{D(m)}] \;\big|\;
  \exists n>m \\
\exists v\!\in\!\dom(\xi)\cap[2^{D(n)}]\setminus B_n: 
   d\big(\xi(u),\xi(v)\big) \!<\! 2^{-m}  \big\} 
\end{multline}
constitutes a name of $W$:
Since the right-hand side of Equation~(\ref{e:Complement})
is r.e., the claim then follows with Remark~\ref{r:Spaces}g).
Indeed, every element $u$ of the right-hand side
arises from some $v\not\in B_n$ with $d\big(\xi(u),x\big) < 2^{-m}$
for $x:=\xi(v)$. $(B_n)$ being a standard name of 
$X\setminus X^\circ$ implies $d_{X\setminus W^\circ}(x)>2^{-n}$
and in particular $x\in W^\circ\subseteq W$.
\\
On the other hand for every $u\in A_m$,
being a standard name implies $d\big(\xi(u),x\big)<2^{-m}$
for some $x\in W=\overline{W^\circ}$.
Hence there exists some $y\in W^\circ$ with 
$d(x,y)<\varepsilon:=2^{-m}-d\big(x,\xi(u)\big)$;
and in turn some integer $n>m$ such that $2^{-n-1}<\varepsilon-d(x,y)$
and $\ball(y,3\cdot2^{-n-1})\subseteq W^\circ$ holds;
and in turn some $v\in\dom(\xi)\cap[2^{D(n)}]$
with $d\big(y,\xi(v)\big)\leq2^{-n-1}$. Then 
$d\big(\xi(v),\xi(u)\big) \;\leq\;$
\[ \leq
\underbrace{d\big(\xi(v),y\big)}_{\leq2^{-n-1}} \;+\;
\underbrace{d\big(y,x\big)}_{<\varepsilon-2^{-n-1}} \;+\; 
\underbrace{d\big(x,\xi(u)\big)}_{=2^{-m}-\varepsilon} \;<\; 2^{-m} \]
and $\ball\big(\xi(v),2\cdot2^{-n-1}\big)\subseteq \ball(y,3\cdot2^{-n-1})\subseteq W^\circ$
implies $d_{X\setminus W^\circ}\big(\xi(v)\big)\geq 2\cdot2^{-n-1}=2^{-n}$
hence $v\not\in B_n$ as the latter is a standard name of $X\setminus W^\circ$.
This demonstrates that every $u\in A_m$ 
is an element of the right-hand side
of Equation~(\ref{e:Complement}).
\item[j)]
Fix a recursive name $(A_m)_{_m}$ of $W$,
a joint recursive time bound and 
binary modulus of continuity $\mu:\IN\to\IN$ of $\Lambda$ according to Theorem~\ref{t:Metric}d),
and uniformly recursive $\Lambda^{\mu}_{n}:A_{\mu(n)}\to\dom(\upsilon)\cap[2^{E(n)}]$
%with $e\Big(\upsilon\big(\Lambda^{\mu}_{n}(a')\big),\Lambda\big(\xi(a')\big)\big)\leq2^{-n}$
according to Remark~\ref{r:Spaces}h). Then the sets
~ $\graph(\Lambda)_m \;:=$
\begin{multline*}
\bigg\{ \bigg( \xi\Big(\Round\big(u,m\big)\Big), 
\upsilon\Big(\Sound\big(\Lambda^\mu_{m+2}(u),m\big)\Big)\bigg) \;\bigg| \\
\bigg| \;u\in A_{\mu(m+2)} \bigg\}  \;\subseteq\; X\times Y
\end{multline*}
are of the form $(\xi\times\upsilon)[C_m]$
for uniformly recursive $C_m\subseteq[D(m)\cdot E(m)]$
and satisfy $\Hausdorff{(d\times e)}\big(\graph(\Lambda),\graph(\Lambda)_m\big)\leq2^{-m}$. 
\\ The converse claim follows from Theorem~\ref{t:Exp}d).
\IEEEQED\end{enumerate}

%%%%%%%%%%%%%%%%%%%%%%%%%%%%%%%%%%%%%%%%%%%%%%%
\subsection{Exponential Objects and Higher-Type Computation} \label{ss:Exp}
This subsection generalizes Theorem~\ref{t:Metric} uniformly,
that is, with $(W,\Lambda)$ not fixed but given as input:
taken from the Cartesian product (Example~\ref{x:Spaces}e)
of the Hausdorff hyper-space $\calK(X)$ over $X$ for $W$,
and for $\Lambda:X\to Y$ from some closed hyper-space 
of \emph{equi}continuous functions to another compact metric space $Y$:
such as to render this new input space in turn 
compact (Fact~\ref{f:Compact}c). The buzzword `hyper' here 
stresses our climbing up the continuous type hierarchy:

\begin{remark} \label{r:Hierarchy}
For $(X,d)$ a compact metric space, and borrowing notation
to hint at the dual of a topological \emph{linear} space,
write $(X',\Supremum{d})$ for the compact hyper-space 
$X':=\Lip_1(X,[0;1])$ of non-expansive real functions.
\begin{enumerate}[\IEEEsetlabelwidth{}]
\item[a)]
If $\diam(X)=1$,
then $X$ embeds isometrically into $X'$ via $\imath:x\mapsto d(x,\cdot)$.
\item[b)]
In this sense, $X$ is a \emph{proper} subset of $X'$
since there exists no isometry from $X'$ to $X$ for reasons of entropy: \\
Consider $Z\subseteq X$ (non-empty and finite but) of maximum cardinality 
such that it holds  $\forall z,z'\in Z: z=z' \vee d(z,z')\geq 1$.
Then every $F:Z\to\{0,1\}$ is 1-Lipschitz; 
and extends to some $\tilde F\in X'$ \cite{Juutinen},
thus having mutual supremum distance $\geq 1$.
This gives rise to $2^{\Card(Z)}>\Card(Z)$ distinct such $\tilde F$:
Mapping them isometrically to $X$ would violate maximality of $Z\subseteq X$.
\item[c)]
On the other hand every compact space, and in particular $X'$, 
is well-known homeomorphic to some compact subset of 
the Hilbert Cube $\prod_{j\in\IN} [0;2^{-j}]=:X$.
So in this topological (rather than metric)
sense $X'$ may actually admit an embedding into $X$.
\item[d)]
For $X=[0;1]^d$, however, $X'$ is not homeomorphic to (a subset of) $X$:
Fix $k\in\IN$ and for $(y_1,\ldots,y_k)\in [0;1]^k$ 
let $f_{\vec y}:[0;1]^d\to[0;1]$
denote the piecewise linear function with 
$f_{\vec y}(j/k,x_2,\ldots,x_d)\equiv y_j/k$. 
Then $\Psi_k:[0;1]^k\ni\vec y\mapsto f_{\vec y}\in X'$ 
is well-defined, injective, and continuous: an embedding.
An embedding $\Phi:X'\to X$ would thus yield 
a continuous injective $\Phi\circ\Psi_k:(0;1)^k\to(0;1)^d$;
contradicting \emph{Invariance of Domain} for $k>d$.
\end{enumerate}
\end{remark}
We now turn compact hyper-space $\calK(X)$ 
into a computably compact metric space, such that
any name of $W\in\calK(X)$ in the sense of Definition~\ref{d:Metric}d)
is the binary encoding of a name of $W\subseteq X$ 
in the sense of Definition~\ref{d:Metric}f), and vice versa:

\begin{definition} \label{d:Exp}
%\begin{enumerate}[\IEEEsetlabelwidth{}]
%\item[a)]
a) For computably compact metric space $(X,d,\xi,D)$,
consider $\big(\calK(X),\Hausdorff{d},\Hausdorff{\xi},2^D\big)$
with
\[ \Hausdorff{\xi}:\; \subseteq\IN \:\ni\:
\sum\nolimits_{j\geq 0} b_j\cdot 2^j \;\mapsto\; \big\{\xi(j):b_j=\sdone\big\}\;\in\; \calK(X) \]
for $b_j\in\{\sdzero,\sdone\}$ in case $\emptyset\neq\{j:b_j=\sdone\}\subseteq\dom(\xi)$, 
$\sum\nolimits_{j\geq 0} b_j\cdot 2^j\not\in\dom\big(\Hausdorff{\xi}\big)$ otherwise.
%\item[b)]
\end{definition}
b) Let $\calC(\subseteq\! X,Y):=\bigcup_{W\in\calK(X)} \calC(W,Y)$
denote the set of partial functions $\Lambda:\subseteq X\to Y$ 
with \emph{compact} domain; similarly for $\calC_\mu(\subseteq\! X,Y)$.
%\end{enumerate}
%
%If $(X,d,\xi,d)$ is $\eta(m)$-separated/rectangular, then so is 
%$\big(\calK(X),\Hausdorff{d},\Hausdorff{\xi},2^D\big)$.\\
c) Consider the continuous embedding
\[ \calC(\subseteq\! X,Y) \;\ni\; \Lambda \;\mapsto \; \graph(\Lambda) 
\;\in\; \calK(X\times Y) \enspace , \]
justified by Fact~\ref{f:Compact}g+h),
by Theorem~\ref{t:Metric}j),
and particularly by Item~e) of the following uniform result:

\begin{theorem} \label{t:Exp}
Let $(X,d,\xi,D)$, $(Y,e,\upsilon,E)$ be computably compact metric spaces
with recursive rounding functions. % $\Round:\dom(\xi)\times\IN\to\dom(\xi)$ 
%and $\Sound:\dom(\upsilon)\times\IN\to\dom(\upsilon)$, respectively.
\begin{enumerate}[\IEEEsetlabelwidth{}]
\item[a)]
The union mapping $\calK(X)\times\calK(X)\ni (V,W)\mapsto V\cup W\in\calK(X)$
is computable.
\item[b)]
The mappings $X\ni x\mapsto\{x\}\in\calK(X)$
and $\calK(X)\supseteq\big\{\{x\}:x\in X\}\ni \{x\}\mapsto x\in X$
are computable.
\item[c)]
There is a computable mapping converting any given name of some
$W\in\calK(X)$ into a standard name of the same $W$.
\item[d)]
For computable $W\in\calK(X)$ and
recursive strictly increasing $\mu:\IN\to\IN$, 
$\graph\big(\calC_\mu(W,Y)\big)$
is a computable compact subset of $\calK(X\times Y)$,
i.e. a computable point in $\calK\big(\calK(X\times Y)\big)$.
\item[e)]
Partial function evaluation is computable, that is, the mapping
%$\calK(X\times Y)\times X \;\supseteq\;$
\begin{multline*}
\calK(X\times Y)\times X \;\supseteq\; \\
\big\{ \big(\graph(\Lambda),x\big) \;\big|\;
    \Lambda\in\calC(W,Y), \; W\in\calK(X), \;x\in W \big\} \\ 
\ni\; \big(\graph(\Lambda),x\big) \;\mapsto\; \Lambda(x) \;\in\; Y \enspace .
\end{multline*}
\item[f)]
The evaluation algorithm from (e) admits a uniformly computable
multivalued runtime bound $T(\Lambda,n)$, i.e.,
depending only on $\Lambda$ and the output precision $n$,
that is simultaneously a binary modulus of continuity:
\begin{multline*}
T: \calK(X\times Y)\times\IN \;\supseteq\; 
\graph\big(\calC(\subseteq\! X,Y)\big)\times\IN \;\ni \\
\ni \; 
\big(\graph(\Lambda),n\big)\;\mapstoto \; m\in\IN : \\
  \forall x,x'\!\in\!\dom(\Lambda): \;
d(x,x')\!>\!2^{-m} \:\vee\:
e\big(\Lambda(x),\Lambda(x')\big)\!\leq\!2^{-n}
\end{multline*}
\item[g)]
Function restriction is computable, i.e. the mapping
\begin{multline*}
\calK(X\times Y)\times\calK(X) \;\supseteq\; \\[0.3ex]
\big\{ \big(\graph(\Lambda),V\big) \;\big|\;
    \Lambda\in\calC(W,Y), \; V,W\in\calK(X), \;V\subseteq W \big\} \\ 
\ni\; \big(\graph(\Lambda),V\big) \;\mapsto\; 
\graph\big(\Lambda\big|_V\big) \;=\; \\
\;=\;\graph(\Lambda)\cap(V\times Y) 
\;\in\; \calK(X\times Y)  \enspace .
\end{multline*}
\item[h)]
Type conversion is also computable: partial evaluation 
\[ \calC(\subseteq\! X\!\times\! Y,Z)\times X \;\ni\; (\Lambda,x)\;\mapsto\; \Lambda(x,\cdot)\;\in\;\calK(\subseteq\! Y,Z) \]
as well as the converse, \emph{un-}`Sch\"{o}nfinkeling' {\rm\cite[p.21]{Strachey}}. 
\vspace*{0.3ex}%
\item[j)]
And so is function image
\[ \calC(\subseteq\! X,Y)\times\calK(X) \;\ni\; (\Lambda,W)\;\mapsto\; \Lambda[W] \;\in\;\calK(Y) \enspace . \]
\item[k)]
Suppose $\Phi:X\to Y$ is computable and \emph{open} in that
images $\Phi[U]\subseteq Y$ of open $U\subseteq X$ are open again.
Then the restricted pre-image mapping
\[ \calKR(Y) \;\ni\; V \;\mapsto\; \Phi^{-1}[V] \;\in\; \calKR(X) \]
is well-defined and computable.
\end{enumerate}
\end{theorem}
Here we denote by $\calKR(X)=\big\{ W\subseteq X : W=\overline{W^\circ}\big\}\subseteq\calK(X)$ 
the family of so-called \emph{regular} subsets of $X$; 
recall Fact~\ref{f:Euclidean}h) and Theorem~\ref{t:Metric}h).

\begin{IEEEproof}[Proof of Theorem~\ref{t:Exp}k)]
Preimage of a continuous open mapping 
commutes with topological closure and interior:
$\Phi^{-1}\big[S^\circ\big]=\big(\Phi^{-1}[S]\big)^\circ$ and 
$\Phi^{-1}\big[\overline{S}\big]=\overline{\Phi^{-1}[S]}$;
cmp. \cite[\textsc{Lemma~4.4}ab]{MLQ2}.
$\Phi^{-1}[V]$ is thus regular.
Moreover both $W:=\Phi^{-1}\big[V\big]$ and 
$\Phi^{-1}\big[Y\setminus V^\circ\big]=Y\setminus W^\circ$
are co-computable according to Theorem~\ref{t:Metric}f);
hence $W$ is computable by virtue of Theorem~\ref{t:Metric}h).
This argument is non-uniform, but closer inspection
shows it to hold uniformly.
\IEEEQED\end{IEEEproof}

%%%%%%%%%%%%%%%%%%%%%%%%%%%%%%%%%%%%%%%%%%%%%%%%%%%%%%%%%%%%%%%
\section[Applications: Fr\'{e}chet Distance and Shape Optimization]{Applications} \label{s:Applications}
We apply the above considerations to two computational problems
over compact metric spaces beyond the classical Euclidean case:
a space of homeomorphisms (Subsection~\ref{ss:Frechet}), 
and the space of compact subsets (Subsection~\ref{ss:Shape}).

%%%%%%%%%%%%%%%%%%%%%%%%%%%%%%%%%%%%%%%%%%%
\subsection{Fr\'{e}chet Distance} \label{ss:Frechet}

In 1906 Maurice Fr\'{e}chet introduced a pseudo-metric for 
parameterized continuous curves and, in 1924, for parameterized surfaces
that in various ways improves over both supremum and Hausdorff Norm:

\begin{definition} \label{d:Frechet}
Let $(X,d)$, $(Y,e)$ be compact metric spaces.
\begin{enumerate}[\IEEEsetlabelwidth{}]
\item[a)]
The \emph{Fr\'echet Distance} of two continuous mappings $A,B:X\to Y$ is given by
$F(A,B) = \inf\nolimits_{\varphi} F_{\id,\varphi}(A,B)$, where
\begin{equation} \label{e:Frechet}
F_{\alpha,\beta}(A,B):=\sup\nolimits_{x\in X} e\Big(A\big(\alpha(x)\big),B\big(\beta(x)\big)\Big)
\end{equation}
with infimum ranging over the set $\Aut(X)$ 
of all homeomorphisms (i.e. continuous bijections) $\varphi:X\to X$. 
\item[b)]
For $X=[0;1]$, the \emph{oriented} Fr\'echet Distance $F'(A,B)$
of continuous (not necessarily simple) curves $A,B:[0;1]\to Y$
is defined similarly with the infimum ranging 
over $\Aut'([0;1])$: the set of all strictly
increasing continuous $\varphi:[0;1]\to[0;1]$
with $\varphi(0)=0$ and $\varphi(1)=1$.
\item[c)] 
For $X=\Sphere^1$ the unit circle,
the \emph{oriented} Fr\'echet Distance $F'(A,B)$
of continuous loops $A,B:\Sphere^1\to Y$
is defined similarly with infimum ranging over $\Aut'(\Sphere^1)$: 
the set of all clockwise continuous bijections $\varphi:\Sphere^1\to\Sphere^1$.
\item[d)]
For $X=\cball^2$ the Euclidean unit disc,
the \emph{oriented} Fr\'echet Distance $F'(A,B)$
of continuous 2D surfaces $A,B:\Sphere^1\to Y$
is defined similarly with infimum ranging over $\Aut'\big(\cball^2\big)$: 
the set of all continuous bijections $\varphi:\cball^2\to\cball^2$
mapping some/all clockwise simple curves in $\cball^2$ to clockwise image(s)
\mycite{Definition~2}{Alt}.
\item[e)]
More generally fix a $d$-dimensional orientable compact manifold $X$,
i.e.,with $d$-th homology group $\Hom_d(X,\IZ)\cong\IZ$
\mycite{Corollary~65.4}{Munkres}.
For any homeomorphism $\varphi:X\to X$,
the action of composition with $\varphi$ induces an isomorphism
of the $k$-th homology group; which for $k=d$ can only be 
multiplication either by $-1$ or by $+1$;
and the latter $\varphi$ by definition comprise $\Aut'(X)$.
\end{enumerate}
\end{definition}
The above notions have recently received much attention ---
in Computational Geometry, that is, for polygonal curves and triangulated surfaces;
cf. for instance \cite{Heekap,Aronov,Driemel,Alt,Godau}
and both the references and motivating examples therein
--- as well as for the important 
\begin{question}
Without restricting to piecewise/combinatorial inputs,
can the Fr\'{e}chet Distance(s) be computed in the sense
of Recursive Analysis, that is, 
by approximation up to guaranteed absolute error $2^{-n}$ 
for every given $n\in\IN$ and every given/fixed 
pair of continuous/computable functions $A,B$ ?
\end{question}
Theorem~\ref{t:Frechet} below gives a positive answer 
for curves ($X=[0;1]$) and loops ($X=\Sphere^1$)
but also shows that an optimal reparametrization 
$\varphi$ cannot in general be computable.

\begin{figure}[htb]
\begin{center}
\includegraphics[width=0.97\columnwidth]{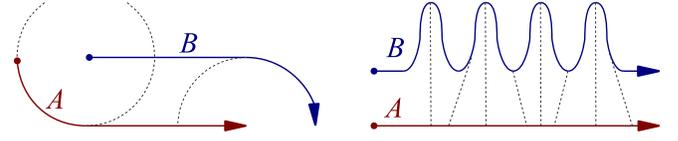}
\caption{\label{f:Frechet}%
a) Two smooth simple curves $A,B:[0;1]\to[0;1]^2$ 
whose Fr\'{e}chet Distance is not attained
by any injective reparameterization $\varphi$.
~ b) Two smooth simple curves $A,B:[0;1]\to[0;1]^2$
whose Fr\'{e}chet Distance is attained
by a continuum of reparameterizations $\varphi$.}
\end{center}
\end{figure}
\begin{figure}[htb]
\begin{center}
\includegraphics[width=0.47\columnwidth]{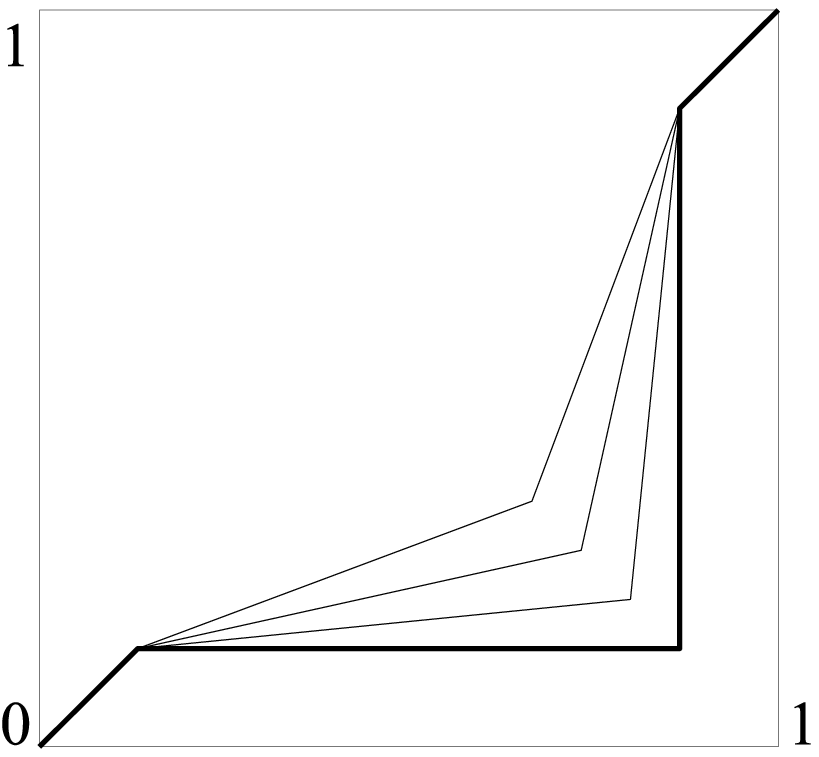}\hspace*{\fill}%
\includegraphics[width=0.46\columnwidth]{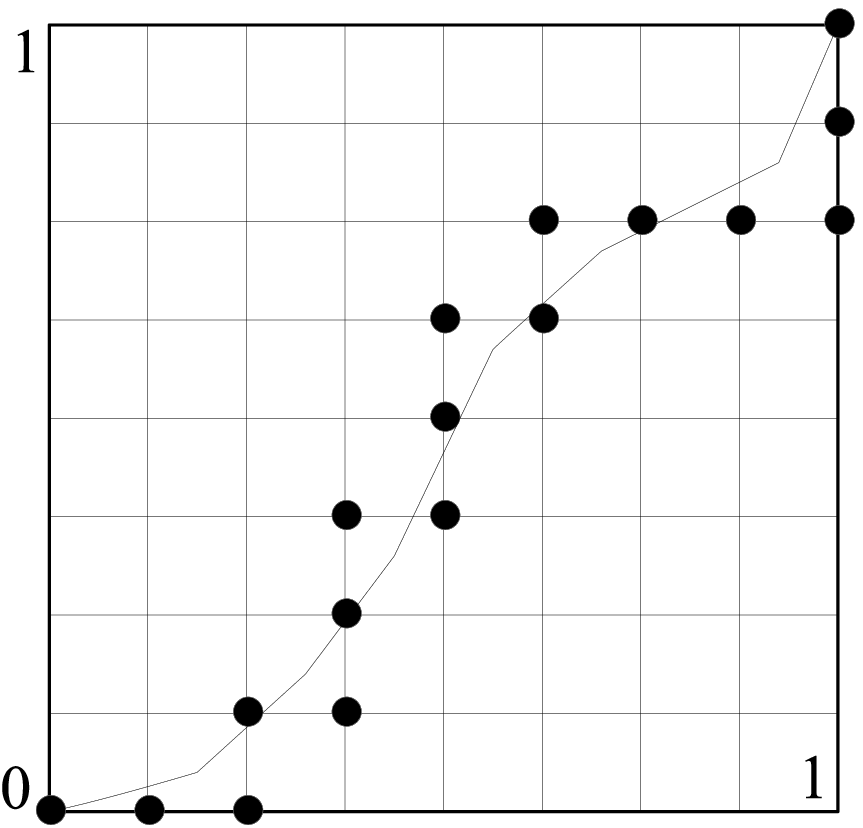}
\caption{\label{f:Graph}%
a) Example of convergence in Hausdorff (graph) but not Supremum (function) norm.
~ b) Illustrating the proof of Theorem~\ref{t:Frechet}a).}
\end{center}
\end{figure}

Recall (Fact~\ref{f:Compact}a) that compactness and continuity
guarantee infimum (e.g. in Definition~\ref{d:Math}f) 
to exist, be attained, and computable
according to Fact~\ref{f:Euclidean}h) and Theorem~\ref{t:Metric}g).
Our goal is to argue similarly in Equation~(\ref{e:Frechet}),
only that the ground space here consists of functions $\varphi$.
A first \naive attempt fails since $\Aut(X)\subseteq\calC(X,X)$
is not compact and the infimum thus not necessarily attained:

\begin{remark} \label{r:Frechet}
\begin{enumerate}[\IEEEsetlabelwidth{}]
\item[a)]
The pseudo-metric in Equation~\ref{e:Frechet} is symmetric:
$\displaystyle F_{\id,\varphi}(A,B) = F_{\alpha\circ\beta,\alpha\circ\varphi\circ\beta}(A,B)$
holds for all bijections $\alpha,\beta:X\to X$, since the set
$\big\{ \big(x,\varphi(x)\big):x \in X\big\}$ agrees with
$\big\{ \big(\alpha\circ\beta(y),\alpha\circ\varphi\circ\beta(y)\big):y \in X\big\}$.
However the $\inf$ over $\varphi\in\Aut(X)$ in Definition~\ref{d:Frechet} 
is in general `attained' only by \emph{non-}injective reparametrizations
(Figure~\ref{f:Frechet}a).
\item[b)]
On the other hand, the mapping  $(\alpha,\varphi)\mapsto F_{\alpha,\varphi}(A,B)$
is uniformly continuous; namely has modulus of continuity the sum of those of $A$ and $B$.
The sets $\Aut(X)$ and $\Aut'(X)$
may thus be replaced by their topological closures,
$\overline{\Aut(X)}$ and $\overline{\Aut'(X)}$
in $\calC(X,X)$ as proper supersets, without
affecting the value of $F$ and $F'$, respectively.
However those closures still lack equicontinuity.
\item[c)]
For the smooth simple curves $A,B:[0;1]\to[0;1]^2$ 
depicted in Figure~\ref{f:Frechet}b), 
their (non-/oriented) Fr\'{e}chet Distance is attained
by a continuum of homeomorphisms $\varphi:[0;1]\to[0;1]$,
i.e., \emph{non-}uniquely.
\item[d)]
$\Aut([0;1])$ is the disjoint union 
of the path-connected subspace of increasing homeomorphisms,
i.e. those in $\Aut'([0;1])$, 
and the decreasing ones; similarly for $\Aut(\Sphere^1)$.
More generally, for any $d$-dimensional orientable compact manifold $X$,
$\Aut(X)$ decomposes into 
the locally arc-connected subspace $\Aut'(X)$ 
of orientation-preserving homeomorphisms
and that of orientation-reversing ones \cite{Sanderson}. 
%Hence in/computability of $F(A,B)$ is equivalent to that of $F'(A,B)$. 
\item[e)] 
To every $\varphi\in\Aut'([0;1])$
there exist 2-Lipschitz $\psi,\chi\in\Aut'([0;1])$  
such that $\varphi=\psi\circ\chi^{-1}$; similarly for the non-oriented case.
%On the other hand to every $L\in\IN$ there exists a $f\in\Aut([0;1]^2)$
%such that no $L$-Lipschitz bijections $g,h:[0;1]^2\to[0;1]^2$ 
%satisfy $f=g\circ h^{-1}$ nor $f=g^{-1}\circ h$.
\item[f)]
There exists a constant $K\geq2$
such that every $\varphi\in\Aut'(\Sphere^1)$
admits a decomposition $\varphi=\psi\circ\chi^{-1}$
with $K$-Lipschitz bijections $\psi,\chi\in\Aut'(\Sphere^1)$;
again, similarly for the non-oriented case.
%where $\Sphere^1=\{x+iy : x^2+y^2=1\}\subseteq\IC$ denotes the 1-sphere.
%but in general no constant $L\in\IN$ can guarantee a 
%decomposition $f=h^{-1}\circ g$ with $L$-Lipschitz bijections $g,h:[0;1]\to[0;1]$.
\item[g)] 
There exists a constant $K\geq2$ such that
every Lipschitz-continuous $\varphi\in\Aut'\big(\cball^d\big)$ 
admits a decomposition $\varphi=\alpha^{-1}\circ\beta\circ\gamma^{-1}\big|_{\cball^d}$
with $K$-Lipschitz $\alpha,\beta,\gamma\in\Aut'\big(2\cball^d\big)$;
similarly for the non-oriented case.
Here, using Minkowski operations, $\cball^d+\cball^d=2\cball^d=\cball(0,2)\subseteq\IR^d$
denotes the closed Euclidean ball around center 0 with radius 2.
%Suppose $A:X\to Y$ has binary modulus of continuity $\mu$,
%$B:X\to Y$ has binary modulus of unicity $\nu$,
%and $\varphi:X\to X$ satisfies $\Supremum{e}(A,B\circ\varphi)=0$.
%Then $\varphi$ has binary modulus of continuity $\mu\circ\nu$.
%Moreover, for continuous/computable $A,B:\cball^d\to Y$,
%their radially constant extensions
%\[ \tilde A: 2\cball^d \;\ni\; \vec x \;\mapsto\;
%\left\{ \begin{array}{cl} A(\vec x) & |\vec x|\leq1 \\
%A(\vec x/|\vec x|) & |\vec x|\geq1 \end{array} \right. \]
%are again continuous/computable and satisfy
%$\Supremum{e}(A,B\circ\varphi)=
%\Supremum{e}\big(\tilde A,\tilde B\circ\beta^{-1}\circ\alpha\big)$.
\item[h)]
Picking up on b), 
extend the definition of $F_{\alpha,\beta}(A,B)$
according to Equation~(\ref{e:Frechet}) from continuous functions 
$\alpha,\beta:X\to X$ to compact relations $\alpha,\beta\subseteq X\times X$
as %~ $F_{\alpha,\beta}(A,B):=$
\[ 
\sup\big\{ e\big(A(a),B(b)\big) \;\big|\;
\exists x\in X: (x,a)\in\alpha, (x,b)\in\beta\big\} \enspace . \]
Then $(\alpha,\varphi)\mapsto F_{\alpha,\varphi}(A,B)$
still remains continuous w.r.t. the Hausdorff metric
on $\calK(X\times X)\times\calK(X\times X)$, and it holds
$F(A,B)=\inf\nolimits_{\varphi} F_{\id,\varphi}(A,B)
 =\inf\nolimits_{\alpha,\beta} F_{\alpha,\beta}(A,B)$
with infimum taken over $\overline{\graph(\Aut(X))}\subseteq\calK(X\times X)$;
similarly for $F'$ and $\Aut'(X)$: see Figure~\ref{f:Graph}a).
\end{enumerate}
\end{remark}
For motivation, consider a bounded uniformly continuous functional
$\Phi:\IP\to\IR$ on non-compact $\IP=\{(\alpha,\varphi):\alpha,\varphi>0\big\}$
but satisfying `scaling invariance' $\Phi(\alpha,\varphi)=\Phi(1,\varphi/\alpha)$.
Then it obviously suffices to consider $(\alpha,\varphi)\in[0;1]^2$:
a compact space. Items~e+f+g) exhibit a similar property for
$(\alpha,\varphi)\mapsto\Supremum{e}(A\circ\alpha,B\circ\varphi)$,
but without commutativity.
\begin{theorem} \label{t:Frechet}
Let $(Y,e,\upsilon,E)$ denote a computably compact space
of $\diam(Y)\leq 1$.
\begin{enumerate}[\IEEEsetlabelwidth{}]
\item[a)]
The compact set 
$\graph\big(\overline{\Lip_2([0;1],[0;1])\cap\Aut'([0;1])}\big)\subseteq\calK([0;1]^2)$ 
of graphs of non-decreasing 2-Lipschitz 
$\varphi:[0;1]\to[0;1]$ with $\varphi(0)=0$ and $\varphi(1)=1$ is computable.
\item[b)]
Non/oriented Fr\'{e}chet Distances between continuous paths
$F,F':\calC([0;1],Y)^2\to[0;1]$
are computable.
\item[c)]
The same holds for non/oriented Fr\'{e}chet Distances between continuous loops
$F,F':\calC(\Sphere^1,Y)^2\to[0;1]$.
\item[d)]
There exist computable smooth $A,B:[0;1]\to[0;1]$ and 
strictly increasing homeomorphism $\varphi:[0;1]\to[0;1]$
such that $A=B\circ\varphi$ holds
but no \emph{computable} non-decreasing surjection $\varphi$
satisfies $A=B\circ\varphi$; 
nor does any computable non-increasing surjection $\varphi$.
\item[e)]
There exist computable smooth \emph{simple} (=injective) $\tilde A,\tilde B:[0;1]\to[0;2]^2$,
codomain considered equipped with the 2D maximum norm, 
such that $F'(\tilde A,\tilde B)=1=F(\tilde A,\tilde B)$ is attained
by some strictly increasing homeomorphism $\varphi:[0;1]\to[0;1]$
but by no \emph{computable} non-decreasing/non-increasing surjection $\varphi$.
\end{enumerate}
\end{theorem}
%
%YYY Figure about closure in function space and graph topology
Regarding higher dimensions, \mycite{Theorem~1}{Buchin} 
has asserted at least left/upper semi-computability;
recall the paragraph following Fact~\ref{f:Euclidean}:
Computationally enumerating a sequence $\varphi_n$
%$:X\to X$ of here piecewise linear homeomorphisms
dense in separable $\Aut(X)\subseteq\calC(X,X)$,
together with computability and continuity of $A,B,\Supremum{e}$
yields a computable sequence $F_{\id,\varphi_n}(A,B)$
whose infimum coincides with $F(A,B)$;
and for $\varphi_n$ ranging over a compact space, its covering property 
%together with uniform continuity of $\varphi\mapsto F_{\id,\varphi}(A,B)$,
asserts that finitely many (balls centered around) them
suffice to approximate $F(A,B)$ also from below.
%but of course these qualitative ideas 
%remain to be made quantitative and computable:

\begin{IEEEproof}[Proof of Theorem~\ref{t:Frechet}]
\begin{enumerate}[\IEEEsetlabelwidth{}]
\item[a)]
Recall that a name of continuous $\varphi:[0;1]\to[0;1]$
is a family of finite sets $C_m\subseteq\ID_m\times\ID_m$
approximating $\graph(\varphi)$ in Hausdorff metric.
Now it is easy to enumerate, uniformly in $m\in\IN$, 
all those $C_m$ satisfying the following condition:
$C_m$ is a `chain' of points in the sense of \emph{Go}
(aka \includegraphics[height=2ex]{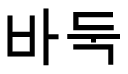}), starting at the lower left
corner and proceeding to the upper right such that at
at least every second step `up' is followed by one `right';
see Figure~\ref{f:Graph}b) illustrating the idea
(that we deliberately refrain from formalizing further).
Then the graph of every $\varphi\in\Lip_2([0;1],[0;1])\cap\Aut'([0;1])$
has Hausdorff distance at most $2^{-m}$ to some such
$C_m$; and conversely every such $C_m$ has distance 
at most $2^{-m}$ to the graph of some 
$\varphi\in\Lip_2([0;1],[0;1])\cap\Aut'([0;1])$.
The collection $\mathfrak{C}_m$ of all those 
$C_m\subseteq\ID_m\times\ID_m$, with $m\in\IN$,
thus constitutes a name of 
$\graph\big(\overline{\Lip_2([0;1],[0;1])\cap\Aut'([0;1])}\big)\subseteq\calK([0;1]^2)$.
\item[b)]
By Remark~\ref{r:Frechet}b+e), $F'(A,B)$ coincides with 
$\inf_{\chi,\psi} F_{\chi,\psi}(A,B)$,
where the infimum ranges over the closet subset
$\overline{\Lip_2([0;1],[0;1])\cap\Aut'([0;1])}\times
\overline{\Lip_2([0;1],[0;1])\cap\Aut'([0;1])}$
of computably compact $\Lip_2([0;1],[0;1])\times\Lip_2([0;1],[0;1])$.
Moreover said subset is computable by a);
and so is the mapping $(\chi,\psi)\mapsto F_{\chi,\psi}(A,B)$ on it.
Hence Theorem~\ref{t:Metric}g) assert its image to be a
computable subset of $[0;1]$, whose minimum is computable
according to Fact~\ref{f:Euclidean}j).
The non-oriented case proceeds similary according to Remark~\ref{r:Frechet}d).
\item[c)]
By Remark~\ref{r:Frechet}b+f) and regarding that
$\overline{\Lip_K(\Sphere^1,\Sphere^1)\cap\Aut'(\Sphere^1)}\times
\overline{\Lip_K(\Sphere^1,\Sphere^1)\cap\Aut'(\Sphere^1)}$
is a computable subset of computably compact
$\Lip_K(\Sphere^1,\Sphere^1)\times\Lip_K(\Sphere^1,\Sphere^1)$
as domain of computable mapping 
$(\chi,\psi)\mapsto F_{\chi,\psi}(A,B)$
The non-oriented case proceeds similary.
\IEEEQED\end{enumerate}\end{IEEEproof}

\begin{IEEEproof}[Proof of Remark~\ref{r:Frechet}]
\begin{enumerate}[\IEEEsetlabelwidth{}]
\item[e)] 
Let $\varphi$ be non-decreasing. Then the continuous 
and surjective mapping $\tilde\varphi:[0;1]\ni t\mapsto \big(t+\varphi(t)\big)/2\in[0;1]$
satisfies, for $t\geq t'$,
\[ \tilde\varphi(t)-\tilde\varphi(t') = (t-t')/2 \;+\; \big(\varphi(t)-\varphi(t')\big)/2 
\;\geq\: (t-t')/2 \]
and hence is strictly increasing with 2-Lipschitz inverse $\chi\in\Aut'([0;1])$.
It remains to 
observe that $\psi:=\varphi\circ\chi$ is 2-Lipschitz since, again for $t\geq t'$,
\begin{multline*}
\psi\big(\tfrac{t+\varphi(t)}{2}\big)
\:-\:
\psi\big(\tfrac{t'+\varphi(t')}{2}\big)
\;=\; \varphi(t) \:-\: \varphi(t')
\;\leq \\
\leq\; \varphi(t) - \varphi(t') + t-t'
\;=\; 2\cdot \big(\tfrac{t+\varphi(t)}{2} - \tfrac{t'+\varphi(t')}{2}\big)
\enspace . \end{multline*}
\item[f)]
Applying an isometric rotation
we may w.l.o.g. suppose $\varphi(1)=1$.
Since $\Sphere^1$ is homeomorphic to $[0;1)\mod 1$,
this reduces to e).
\item[g)]
According to f),
the restriction of $\varphi$ to the ball's boundary
admits a decomposition $\varphi\big|_{\Sphere^1}=\psi\circ\chi^{-1}$
with $K$-Lipschitz homeomorphisms $\psi,\chi:\Sphere^1\to\Sphere^1$.
Note that $\vec x\mapsto \vec x/|\vec x|_2$ is $L$-Lipschitz 
outside the Euclidean disk $\ball^d(0,1/L)$ of radius $1/L$.
Thus, applying \emph{Alexander's Trick}, 
$\chi$ extends radially to a $2K$-Lipschitz homeomorphism 
of entire $\cball^2$, and of $2\cball^2$, via
~ $\chi(\vec x) \;:=\; \chi(\vec x/|\vec x|_2)\cdot|\vec x|$. ~
So $\varphi\circ\chi\big|_{\cball^2}\in\Aut(\cball^2)$ 
coincides on $\Sphere^1$ with $\psi$ and in particular is $2K$-Lipschitz there.
Now abbreviate $L:=2K$ and define
\begin{gather*}
\beta:\vec x\mapsto\;\left\{ \begin{array}{c@{\;:\;}l}
\varphi\circ\chi(\vec x)/L & |\vec x|_2\leq1 \\
b_L(|\vec x|_2)\cdot \varphi\circ\chi(\vec x/|\vec x|_2) & |\vec x|_2\geq 1
\end{array}\right., \\[0.5ex]
\tilde\alpha:\vec y\mapsto\;\left\{ \begin{array}{c@{\;:\;}l}
\vec y \cdot L & |\vec y|_2\leq1/L \\
a_L(|\vec y|_2)\cdot\vec y/|\vec y|_2 & |\vec y|_2\geq 1/L
\end{array}\right.
\end{gather*}
for affine $b_L(r):=(2-1/L)\cdot r-2+2/L$
and $a_L(s):=\tfrac{L\cdot s+2L-2}{2L-2}$; cmp. Figure~\ref{f:Alexander}.
Then, $b_L:\big[1;2\big]\to\big[\tfrac{1}{L};2\big]$ 
constituting an increasing bijection,
implies that $\beta:2\cball^d\to2\cball^d$ is well-defined,
continuous, injective, surjective, and $2L$-Lipschitz.
Similarly, $a_L:\big[\tfrac{1}{L};2\big]\to\big[1;2\big]$
constituting an increasing bijection, 
implies that $\tilde\alpha:2\cball^d\to2\cball^d$
is well-defined, continuous, injective, and surjective 
with $2L$-Lipschitz inverse $\alpha:=\tilde\alpha^{-1}$ 
since $a_L\circ b_L=\id_{[1;2]}$.
Finally, $\varphi\circ\chi=\tilde\alpha\circ\beta\big|_{\cball^d}$
is easily verified.
\IEEEQED\end{enumerate}\end{IEEEproof}

%%%%%%%%%%%%%%%%%%%%%%%%%%%%%%%%%%%%%%%%%%%
\subsection{Shape Optimization} \label{ss:Shape}
This subsection exhibits weak conditions that assert
computability of the following
generic optimization problem:

\begin{definition} \label{d:Opt}
For fixed $X$ and 
given $\Lambda,\Phi:X\to\IR$, 
determine the real number
$r:=\max\Lambda\Big[\Phi^{-1}\big[(-\infty;0]\big]\Big]$,
provided it exists.
$\Lambda$ is the \emph{objective function},
$r$ its \emph{optimum} 
w.r.t. \emph{constraint} $\Phi\leq0$,
the latter called \emph{feasible} if $0\in\Phi[X]$.
\end{definition}

\medskip\medskip
To guarantee said existence, it suffices
that (i) $X$ be compact, 
(ii) $\Lambda$ be continuous,
and (iii) $\Phi$ be lower semi-continuous:
Condition (iii) asserts that the non-empty
domain $\Phi^{-1}\big[(-\infty;0]\big]\subseteq X$
is closed and thus compact by (i), 
hence (ii) $\Lambda$ attains its maximum on it.
Conversely simple counterexamples
show that none of the three conditions
can in general be omitted.

\begin{example} \label{x:Opt}
\begin{enumerate}[\IEEEsetlabelwidth{}]
\item[a)]
Linear Optimization refers, up to scaling,
to the case $X=[0;1]^d$ with $\Lambda:X\to\IR$ linear 
together with a finite conjunction of (w.l.o.g. non-constant) linear constraints
$\Phi_j:\vec x\mapsto b_j+\vec x\cdot\vec a_j^+$, $\vec a_j\neq0$,
collected into the single $\Phi:=\max_{j} \Phi_j$. 
Note that $\Lambda$ and $\Phi$ are continuous,
and $\Phi$ is furthermore open.
\item[b)]
Convex Optimization refers, again up to scaling,
to the case $X=[0;1]^d$ but now permits the generalization
to convex negated objective functions $-\Lambda:X\to\IR$ and 
non-constant constraints $\Phi_j$, again subsumed in the
(still convex) single $\max_{j} \Phi_j$.
\item[c)]
In Discrete Optimization, $X$ is a finite but `large'
set. Equipped with the discrete topology, 
it renders every $\Lambda,\Phi:X\to\IR$ continuous.
\item[d)]
\emph{Shape Optimization} refers, again up to scaling,
to the case $X=\calK([0;1]^d)$ with
$\Lambda,\Phi\in\calC(X)$; cmp. \cite{Pierre}.
\item[e)]
The (Lebesgues) measure, considered as mapping
$\Vol_1:\calKR([0;1])\to[0;1]$, is upper semi-continuous 
but not continuous nor computable.
\end{enumerate}
\end{example}
\medskip
Recall $\calKR(X)=\big\{ W\subseteq X : W=\overline{W^\circ}\big\}$.
For a fixed convex compact subset $X$ of a Fr\'{e}chet (i.e. complete translation invariant metric vector) space,
write $\calKC(X)\subseteq\calK(X)$ for the family of its convex compact non-empty subsets.
%and $\calKCR(X)\subseteq\calKC(X)$ 
%for those convex compact subsets having non-empty interior.

Shape Optimization has recently grown a hot topic in Numerical Engineering \cite{Sokolowski} 
but generally lacks mathematical specification and rigorous algorithmic analysis.
We establish that this, as well as the generic 
optimization problem from Definition~\ref{d:Opt},
is computable for (i') computably compact metric 
space $X$, (ii') computable objective function
and (iii') open computable constraint:

\begin{theorem} \label{t:Opt}
Let $(X,d,\xi,D)$ and $(Y,e,\upsilon,E)$
denote computably compact metric spaces.
\begin{enumerate}[\IEEEsetlabelwidth{}]
\item[a)]
For $X$ any convex compactly computable subset of Euclidean space
with non-empty interior,
$\calKC(X)\subseteq\calK(X)$ is in turn computably compact.
\item[b)]
The generic optimization problem
\begin{multline*}
\calC(X,[0;1])\times\big(\calC(X,[-1;1])\cap\calO_0(X)\big) \;\ni \\
\ni\; (\Lambda,\Phi) \;\mapsto\; 
\max\Lambda\Big[\Phi^{-1}\big[[-1;0]\big]\Big] \;\in\; [0;1]
\end{multline*}
is computable, where $\calO_0(X)$ denotes the family 
of open $\Phi:X\to\IR$ with $0\in\Phi[X]$.
\item[c)]
As opposed to Example~\ref{x:Opt}e), the mapping
$\Vol_d:\calKC\big([0;1]^d\big)\to[0;1]$ is (continuous and) computable.
\item[d)] 
The mapping $\calKC\big([0;1]^d\big)\ni W\mapsto \Area_{d-1}(\partial W)\in[0;2d]$
is well-defined, open, (continuous and) computable, 
where $\Area_{d-1}(\partial W)$
denotes the area measure of $W$'s boundary.
\end{enumerate}
\end{theorem}
Claim~a) is a minor strengthening of \emph{Blaschke's Selection Theorem};
b) follows by 
combining Theorem~\ref{t:Exp}j+k) with Fact~\ref{f:Euclidean}j)
since $[-1;0]$ is regular.

\medskip
As an example `application', consider the
classical \emph{Isoperimetric Problem}
%in $d$-dimensional Euclidean space:
asking to maximize $\Vol_d(W)$ subject to
the constraint $\Area(\partial W)-1\leq 0$, say.
Combining the items of Theorem~\ref{t:Opt},
we conclude that the solution is computable!

On second thought this comes at no surprise, though:
Knowing that the optimal shape is a Euclidean ball,
the solution is easily calculated explicitly
as $1/(4\pi)$ in dimension 2,
$1/(6\sqrt{\pi})$ in dimension 3,
and similar expressions can be derived
in any dimension $d$ involving the gamma function
at half-integral arguments.

%%%%%%%%%%%%%%%%%%%%%%%%%%%%%%%%%%%%%%%%%%%%%%%%%%%%%%%%%%%%%%%
\section{Conclusion and Perspectives} \label{s:Conclusion}
For computably compact metric spaces $(X,d,\xi,D)$ and $(Y,e,\upsilon,E)$
in the sense of Definition~\ref{d:Metric}, we have (i) turned the 
hyper-space $\calK(X)$ of non-empty compact subsets of $X$ into 
computably compact metric space, again; and (ii) similarly for
the space $\calC_\mu(W,Y)$ of partial equicontinuous functions 
$\Lambda:W\to Y$ having non-empty compact domain $W\subseteq X$.
The latter proceeds by identifying such $\Lambda$ 
with $\graph(\Lambda)\in\calK(X\times Y)$; and was shown 
to render evaluation uniformly computable. This generalizes
well-known results for the Euclidean, to arbitrary compact
metric, spaces -- including a hierarchy of higher types.

\medskip\noindent
\subsection*{Perspectives}
\begin{enumerate}[\IEEEsetlabelwidth{}]
\item[a)] Definition~\ref{d:Metric}b) employs dense enumerations,
 that is, surjective partial mappings from $\IN$. When 
 categorically constructing enumerations of the Cartesian product
 (Example~\ref{x:Spaces}d) and Hausdorff hyperspace 
 (Example~\ref{x:Spaces}e), we thus had to `flatten' back 
 the domain of the enumeration: which works from the perspective
 of computability, but is unnatural and complexity-theoretically
 superseded \cite{AkiSTOC}. Future work will instead choose
 a suitable axiomatized category of discrete ground spaces.
\item[b)]
 Speaking of complexity, 
 we will refine the above computability
 investigations to quantitative upper and (using
 adversary arguments in the bit-cost model) lower
 complexity bounds in terms of the separation parameter  $\eta$
 from Definition~\ref{d:Metric};
 cmp. \cite{MuellerZhao,AkiSTOC,Roesnick,metric,Steinberg}.
\item[c)]
 Encoding continuous functions via their graphs generalizes
 to closed relations aka multi-(valued)functions \cite{Henkin}
 known essential in computing on continuous data.
 Regarding Remark~\ref{r:Hierarchy}c+d),
 we don't know whether $[0;1]''$ is homeomorphic
 to (a subset) of $[0;1]'$.
\item[d)]
 Computability of the 2D Fr\'{e}chet Distance remains
 elusive --- until we can establish computability of the compact set
 $\overline{\big\{\graph\big(\Aut([0;1]^2)\big)\big\}}\subseteq\calK\big([0;1]^4\big)$;
 recall Remark~\ref{r:Frechet}h). This boils down to the following question:
\item[e)]
 Given a subset of $\ID_n^4$, (how) can we decide
 whether it has Hausdorff distance
 $\leq2^{-n}$ to the graph of any $\varphi\in\Aut([0;1]^2)$ ?
\end{enumerate}
%\end{perspectives}

\noindent%\textsc{Acknowledgement:}
We thank Akitoshi Kawamura, 
Matthias Schr\"oder,
and Florian Steinberg
for seminal discussions.

%%%%%%%%%%%%%%%%%%%%%%%%%%%%%%%%%%%%%%%%%%%%%%%%%%%%%%%%%%%%%%
%\cleardoublepage
%\IEEEtriggercmd{\itemsep5pt}

%%%%%%%%%%%%%%%%%%%%%%%%%%%%%%%%%%%%%%%%%%%%%%%%%%%%%%%%%%%%%%%%%%%%%%
\pagebreak
\appendix
\begin{IEEEproof}[Proof of Fact~\ref{f:Compact}]
\begin{enumerate}[\IEEEsetlabelwidth{}]
\item[a)]
See \mycite{Theorems~3.11+2.35+4.16}{Rudin}.
\item[b)]
See \mycite{Theorems~4.14+4.19}{Rudin}.
\item[c)]
See \mycite{Theorem~7.25}{Rudin}, 
cmp. also \cite[\textsc{Lemma~13}d]{metric}.
\item[e)]
See \mycite{Theorem~4.26}{Kechris}.
\item[f)]
$x_{\xi,D}$ is obviously closed in Baire space;
and so is $W_{\xi,D}$: If $x_k$ is a sequence in $W$
and $\bar u^{(k)}\in x_{k,\xi,D}$ 
converges to some $\bar u^{(0)}\in\IN^\IN$, 
then triangle inequality yields
\[ d(x_k,x_\ell) \leq d\big(x_k,\xi(u_m^{(k)})\big) +
d\big(\xi(u_m^{(k)}),x_\ell\big) \leq 2^{-m}+2^{-m} \]
for all $\ell,k$ so large such that 
$\beta\big(\bar u^{(k)},\bar u^{(\ell)}\big)<2^{-m}$
hence $u_m^{(k)}=u_m^{(\ell)}=u_m^{(0)}$.
Thus $(x_k)$ constitutes a Cauchy sequence in compact $W$, 
converging by (a) to some $x_0\in W$.
To see $\bar u^{(0)}\in x_{0,\xi,D}$, observe
\[ 2^{-m} \geq d\big(x_k,\xi(u_m^{(k)})\big) =
d\big(x_k,\xi(u_m^{(0)})\big) \overset{k\to\infty}{\longrightarrow}
d\big(x_0,\xi(u_m^{(0)})\big) \]
Regarding relative compactness, 
record that the sets $x_{\xi,D,m}$ are finite, 
in fact of cardinality bounded by $D(m)$ independently of $x\in X$.
Therefore an inital segment $(u_0,\ldots,u_{m-1})$ of any $\bar u\in x_{\xi,D}$ 
can have no more than $D(m)$ possible successors $u_m$.
This asserts that the set $X_{\xi,D}^*$
of finite initial segments of elements in $X_{\xi,D}$ 
constitute a finitely branching subtree of $\IN^*$.
Now apply (d).
\item[g)]
Triangle inequality of $d\times e$ and compactness of $X\times Y$ are immediate.
If $f$ is continuous and $\big(x_n,f(x_n)\big)$ a sequence in $\graph(f)$,
then compactness of $X$ yields a convergent subsequence $x_{n_k}\to x\in X$,
and $f(x_{n_k})\to f(x)$ by continuity. Conversely, the
pre-image $f^{-1}[V]$ of any closed $V\subseteq Y$
coincides with $\pi_1\big[ (X\times V)\cap\graph(f) \big]\subseteq X$
and thus is closed,
where $\pi_1(x,y):=x$ denotes the continuous 
and closed projection map and 
$(X\times V)\cap\graph(f)\subseteq X\times Y$ is closed.
\item[h)]
For every $\big(x,f(x)\big)\in\graph(f)$, 
$\big(x,g(x)\big)\in\graph(g)$ has
$(d\times e)$-distance 
$e\big(f(x),g(x)\big)\leq\Supremum{e}(f,g)$.
Conversely, to every $x\in X$ there exists some $x'\in X$ such that 
$d(x,x'),e\big(f(x),g(x')\big)\leq \Hausdorff{(d\times e)}\big(\graph(f),\graph(g)\big)$;
hence, for $\omega$ a modulus of continuity of $g$,
\begin{multline*}
e\big(f(x),g(x)\big) \;\leq \\
\leq\;
e\big(f(x),g(x')\big) \;+\; e\big(g(x'),g(x)\big) \;\leq\\
\leq\; \Hausdorff{(d\times e)}\big(\graph(f),\graph(g)\big) 
\;+\; \omega\big(d(x',x)\big) \enspace .
\end{multline*}
In particular, uniform convergence $\calF\ni f_n\to f$ 
implies $\graph(f_n)\to\graph(f)$ in Hausdorff distance;
and $\graph(f_n)\to\graph(f)$ with $f\in\calC(X,Y)$
implies $f_n\to f$ in uniform norm according to b).
\IEEEQED\end{enumerate}\end{IEEEproof}

\noindent
\begin{IEEEproof}[Proof of Remark~\ref{r:Frechet}b)]
\begin{eqnarray*}
\lefteqn{|\Supremum{e}(A\circ\alpha,B\circ\varphi)
-\Supremum{e}(A\circ\tilde\alpha,B\circ\tilde\varphi)|\;\leq} \\
&\leq& 
|\Supremum{e}(A\circ\alpha,B\circ\varphi)-
\Supremum{e}(A\circ\alpha,B\circ\tilde\varphi)|
\;+\; \\
&+& 
|\Supremum{e}(A\circ\alpha,B\circ\tilde\varphi)
-\Supremum{e}(A\circ\tilde\alpha,B\circ\tilde\varphi)| \\
&\leq&
\Supremum{e}(B\circ\varphi,B\circ\tilde\varphi)
+ \Supremum{e}(A\circ\alpha,A\circ\tilde\alpha) \\
&\leq& \omega_B\big(\Supremum{e}(\varphi,\tilde\varphi)\big)
+\omega_A\big(\Supremum{e}(\alpha,\tilde\alpha)\big)
\end{eqnarray*}
by reverse triangle inequality for moduli of continuity
$\omega_A,\omega_B$ of $A,B$ respectively.
\IEEEQED\end{IEEEproof}

\noindent
\begin{IEEEproof}[Proof of Theorem~\ref{t:Exp}]
\begin{enumerate}[\IEEEsetlabelwidth{}]
\item[a)] similarly to the proof of Theorem~\ref{t:Metric}a).
\item[b)] similarly to the proof of Theorem~\ref{t:Metric}b+c).
\item[c)] similarly to the proof of Remark~\ref{r:Spaces}f).
\item[d)] similarly to the proof of Theorem~\ref{t:Metric}j).
\item[e)]
Let $(u_m)$ with $u_m\in\dom(\xi)\cap[2^{D(m)}]$ be a given name of $x\in W$
and $(C_m)$ with 
$C_m\subseteq\big(\dom(\xi)\cap[2^{D(m)}]\big)\times\big(\dom(\upsilon)\cap[2^{E(m)}]\big)$
one of $\graph(\Lambda)\subseteq X\times Y$
for some continuous $\Lambda:W\to Y$.
For $n\in\IN$, search for, and output, some 
$v_n\in\dom(\upsilon)\cap[2^{E(n)}]$ 
such that there exists an $m\in\IN$ satisfying the following:
\begin{multline} \label{e:Eval}
\forall (u',v')\in C_m: \quad
d\big(\xi(u_m),\xi(u')\big)>2^{-m+1} \;\vee \\
\vee\; 
e\big(\upsilon(v_n),\upsilon(v')\big) < 2^{-n} \!-\!2^{-m} \enspace . 
\end{multline}
First, Property~(\ref{e:Eval}) is obviously r.e..
Second, any $v_n$ found satisfies
$e\big(\upsilon(v_n),y\big)<2^{-n}$ for $y:=\Lambda(x)$
with $x\in W$ s.t. $d\big(\xi(u_m),x\big)\leq2^{-m}$:
$\Hausdorff{(d\times e)}\big(\graph(\Lambda),C_m\big)\leq2^{-m}$
with $(x,y)\in\graph(\Lambda)$ yields
some $(u',v')\in C_m$ with $d\big(\xi(u'),x\big),e\big(\upsilon(v'),y\big)\leq2^{-m}$;
hence $d\big(\xi(u_m),\xi(u')\big)\leq2^{-m+1}$ by triangle inequality,
so the second part of Property~(\ref{e:Eval}) applies.
Third, there exists some $v_n\in\dom(\upsilon)\cap[2^{E(n)}]$
with $e\big(\upsilon(v_n),y\big)\leq2^{-n-1}$; and,
as $\Lambda$ is defined and continuous at $x$, 
there is some $m\geq n+3$ such that every $x'\in W$ 
with $d(x,x')\leq2^{-m+2}$
satisfies $e\big(y,\Lambda(x')\big)\leq2^{-n-2}$.
In particular every $(u',v')\in C_m$
with $d\big(\xi(u_m),\xi(u')\big)\leq2^{-m+1}$,
having $d\big(\xi(u'),x'\big),e\big(\xi(v'),\Lambda(x')\big)\leq2^{-m}$
for some $x'\in W$ implies $e\big(\upsilon(v_n),\upsilon(v')\big)\leq$
\begin{multline*}
e\big(\upsilon(v_n),y\big)\;+\;e\big(y,\Lambda(x')\big)\;+\;e\big(\Lambda(x'),v'\big)
\;\leq\\
\leq\; 2^{-n-1}+2^{-n-2}+2^{-m}\;\leq\; 2^{-n}-2^{-m} 
\enspace . \end{multline*}
\item[f)]
Observe that Theorem~\ref{t:Metric}d+e) relativizes:
For any oracle $\Oracle$, if $W\in\calK(X)$ is
(co-)computable with $\Oracle$ and $\Lambda:W\to Y$ 
is computable with $\Oracle$, then $\Lambda$
has a runtime bound/binary modulus of continuity
$T(\Lambda,\cdot):\IN\to\IN$ computable with $\Oracle$.
Now for $\Oracle$ encoding names of $W$ and $\Lambda$,
any query made by the oracle machine computing said $T(\Lambda,\cdot)$
can be answered by performing a look-up on the input
$\graph(\Lambda)$. 
For instance the projection (of a name of)
$\graph(\Lambda)\subseteq X\times Y$
onto $X$ yields (a name of) $\dom(\Lambda)=W$.
\item[g)]
By c) w.l.o.g. let $(A_m)$ %with $A_m\subseteq\dom(\xi)\cap[2^{D(m)}]$
denote a given \emph{standard} name of $V\subseteq W$ and $(B_m)$ 
%with $B_m\subseteq\big(\dom(\xi)\cap[2^{D(m)}]\big)\times\big(\dom(\upsilon)\cap[2^{E(m)}]\big)$
one of $\graph(\Lambda)\subseteq X\times Y$,
where $\Lambda:W\to Y$ has binary modulus of continuity $\mu$ according to f).
Abbreviate $n:=\mu(m+2)+1\geq m+2$ and verify 
\[ 
C_m \;:=\; \big\{ \big(\Round(u,m),\Sound(v,m)\big) \;\big|\;
(u,v)\in B_{n}, \; u\in A_{n} \big\}  \]
constituting a name of $\graph(\Lambda|_V)$:
To every $(x,y)\in\graph(\Lambda|_V)$ there exist
$u\in\dom(\xi)\cap[2^{D(n)}]$  and
$v\in\dom(\upsilon)\cap[2^{E(n)}]$ with
$d\big(x,\xi(u)\big),e\big(y,\upsilon(v)\big)\leq2^{-n-1}\leq2^{-m-1}$;
hence $u\in A_n$ and $(u,v)\in B_n$: since these are standard names.
Then $u':=\Round(u,m)$ and $v':=\Sound(v,m)$ with $(u',v')\in C_m$ have 
$d\big(\xi(u'),\xi(u)\big),e\big(\upsilon(v'),\upsilon(v)\big)\leq2^{-m-1}$,
so $(d\times e)\big((\xi(u'),\upsilon(v')),(x,y)\big)\leq2^{-m}$.
\\
Conversely, to every $(u',v')\in C_m$ there exist
by definition $(u,v)\in B_n$ with $u\in A_n$ such that 
$d\big(\xi(u'),\xi(u)\big),e\big(\upsilon(v'),\upsilon(v)\big)\leq2^{-m-1}$;
and $(x,y)\in\graph(\Lambda)$ as well as $x'\in V$ with
$d\big(x',\xi(u)\big),d\big(x,\xi(u)\big),e\big(y,\upsilon(v)\big)\leq2^{-n}$.
Hence $d(x,x')\leq2^{-n+1}=2^{-\mu(m)}$ implies $e(y',y)\leq2^{-m-2}$
for $y':=\Lambda(x')$; and therefore it holds
$d\big(x',\xi(u')\big)\leq2^{-n}+2^{-m-1}\leq2^{-m}$ as well as
$e\big(y',\upsilon(v')\big)\leq2^{-m-2}+2^{-n}+2^{-m-1}\leq2^{-m}$.
\item[h)]
Similar to g) and Theorem~\ref{t:Metric}j), respectively.
\item[j)]
Similar to Theorem~\ref{t:Metric}g).
\IEEEQED\end{enumerate}\end{IEEEproof}

\begin{figure}[htb]
\begin{center}
\includegraphics[width=0.99\columnwidth]{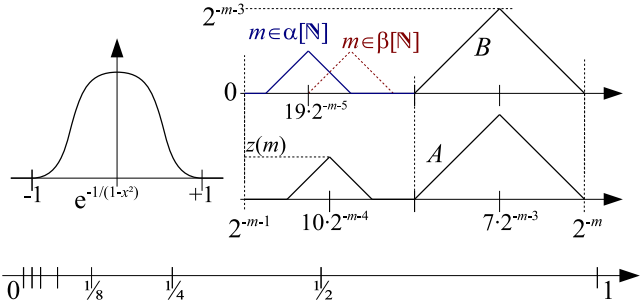}%
\caption{\label{f:Inseparable}%
Illustrating the proof Theorem~\ref{t:Frechet}d)}
\end{center}
\end{figure}

\noindent
\begin{IEEEproof}[Proof of Theorem~\ref{t:Frechet}]
\begin{enumerate}[\IEEEsetlabelwidth{}]
\item[d)]
Let $\alpha,\beta:\IN\to\IN$ denote total recursive injective enumerations
of recursively inseparable sets $\alpha[\IN],\beta[\IN]\subseteq\IN$;
cmp. \mycite{Exercise~1.6.26}{Soare}.
Intuitively, although inputs $m\not\in\alpha[\IN]\cup\beta[\IN]$
may be accepted or rejected arbitrarily, provably
no total algorithm can make such a decision.
\\
Now for each $m\in\IN$ abbreviate $z(m):=0$ 
for $m\not\in\alpha[\IN]\uplus\beta[\IN]$ and
$z(m):=2^{-m^2-\min\{k:\alpha(k)=m\vee\beta(k)=m\}}$ otherwise:
a computable sequence of real numbers;
and so is $x(m)$, defined as $z(m)$ if $m\in\alpha[\IN]$ and $x(m):=0$ otherwise;
and $y(m):=z(m)$ if $m\in\beta[\IN]$ and $y(m):=0$ otherwise.
Next consider the computable 1-Lipschitz `hat' function
$\psi(x)=\max\big\{0,1-|x|\}$
or the computable smooth `pulse' function
$\tilde\psi(x)=\exp\big(-\tfrac{1}{1-x^2}\big)$
for $|x|\leq1$, $\tilde\psi(x):\equiv0$ for $|x|\geq1$:
both have support $[-1;1]$, see Figure~\ref{f:Inseparable}.
Finally define $A(x)$ and $B(x)$, respectively, as
\begin{gather*}
\sum\limits_m \tilde\psi\big(2^{m+3}\cdot x-7\big)/2^{m+3} \;+\;
     \tilde\psi\big(2^{m+4}\cdot x-10\big)\cdot z(m), \\
\sum\limits_m \tilde\psi\big(2^{m+3}\cdot x-7\big)/2^{m+3} \;+\;
     \tilde\psi\big(2^{m+5}\cdot x-19\big)\cdot x(m) \\[-1ex]
\hspace*{25ex} \;+\;
     \tilde\psi\big(2^{m+5}\cdot x-21\big)\cdot y(m)
\end{gather*}
and note that all terms in each sum have disjoint supports.
Moreover, the thus well-defined $A,B:[0;1]\to[0;1]$ are continuous
(particularly at 0) and even smooth, since the $d$-th derivative 
of the $m$-th term is bounded by $z_m\cdot(2^{m+4})^d\leq 2^{-m^2+dm+4d}\to0$
as $m\to\infty$.
Next, it holds $A=B\circ\varphi$ for the
increasing continuous surjection $\varphi:[0;1]\to[0;1]$
mapping the closed interval 
$\big[6\cdot2^{-m-3};11\cdot8\cdot2^{-m-3}\big]$ isometrically to
$\big[6\cdot2^{-m-3};11\cdot8\cdot2^{-m-3}\big]$ 
to match the `large' pulses as well as the `small' ones by mapping
$\big[9\cdot2^{-m-4};11\cdot2^{-m-4}\big]$ 
to $\big[17\cdot2^{-m-5};21\cdot2^{-m-5}\big]$ in case $m\in\alpha[\IN]$
and to $\big[19\cdot2^{-m-5};23\cdot2^{-m-5}\big]$ in case $m\in\beta[\IN]$;
arbitrarily in case $m\not\in\alpha[\IN]\cup\beta[\IN]$.
Conversely, no non-increasing surjective $\varphi$
can make the large pulses match nor satisfy $B=A\circ\varphi$;
and any non-decreasing $\varphi$ that does,
must necessarily map $10\cdot2^{-m-4}$ to $19\cdot2^{-m-5}$
in case $m\in\alpha[\IN]$, to $21\cdot2^{-m-5}$
in case $m\in\beta[\IN]$, and anywhere in
$\big[8\cdot2^{-m-4};12\cdot2^{-m-4}\big]$
in case $m\not\in\alpha[\IN]\cup\beta[\IN]$.
A computable such (total!) $\varphi$ would allow for a 
recursive separation of $\alpha[\IN]$ from $\beta[\IN]$:
Given $m\in\IN$, accept if the $2^{-m-6}$-approximation
to $\varphi(10\cdot2^{-m-4})$ is less than 
$20\cdot2^{-m-5}$ and reject otherwise:
Contradiction.
\item[e)]
Continuing d), consider computable injective
$\tilde A:[0;1]\ni x\mapsto\big(x,A(x)\big)\in[0;1]^2$
and $\tilde B(x):=\big(x,B(x)+1\big)\in[0;2]^2$:
The offset in $y$-coordinate dominates their pointwise
distance in the 2D maximum norm over 
local reparametrization in $x$-coordiate.
\IEEEQED\end{enumerate}\end{IEEEproof}

\begin{figure}[htb]
\begin{center}
\includegraphics[width=0.97\columnwidth]{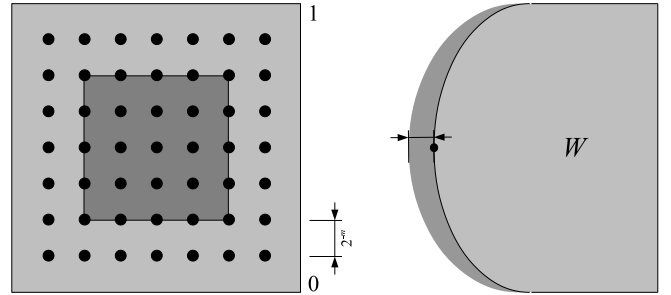}%
\caption{\label{f:Convex}%
$d$-dimensional volume of convex sets is $2d$-Lipschitz
in the worst-case, (unscaled) boundary area is $4d\cdot(d-1)$-Lipschitz
and open a function.}
\end{center}
\end{figure}

\noindent
\begin{IEEEproof}[Proof of Theorem~\ref{t:Opt}]
\begin{enumerate}[\IEEEsetlabelwidth{}]
\item[c)]
$d$-dimensional volume of convex subsets of $[0;1]^d$
is $2d$-Lipschitz continuous, worst-case depicted 
in Figure~\ref{f:Convex}:
The unit hypercube has volume 1, 
shrinking it by $\varepsilon:=2^{-m}$ on each side
yields volume $(1-2\varepsilon)^d\approx 1-2d\varepsilon$.
Thus, for $C_m\subseteq\ID_m^d$ approximating $W\in\calKC([0;1]^d)$
up to error $2^{-m}$ in Hausdorff distance,
the volume of the convex hull of $C_m$ approximates 
the volume of $W$ up to error $2d\cdot 2^{-m}$.
\item[d)] 
Slighly `pulling' any extreme point of convex compact $W\subseteq[0;1]^d$ 
in/outwards, yields a proper sub/superset $W'$ which is again convex compact
but with area of the boundary strictly larger/smaller than that of $W$,
cmp. \myurl{math.stackexchange.com/questions/262568}:
This shows that surface area measure is open a mapping;
cmp. Figure~\ref{f:Convex}.
The latter also depicts a worst-case to $4d(d-1)$-Lipschitz continuity of 
$\calKC([0;1]^d)\ni W\mapsto$
\[ \mapsto\Area_{d-1}(\partial W)
=\liminf\limits_{\delta\to0} 
\frac{\Vol_d(W+\delta\cdot\cball^d)-\Vol_d(W)}{\delta} \]
according to Minkowski-Steiner:
The unit hypercube has surface area $2d$,
shrinking it by $\varepsilon:=2^{-m}$ 
yields surface area $2d\cdot (1-2\varepsilon)^{d-1}\approx 2d-4d\cdot(d-1)$.
Thus, for $C_m\subseteq\ID_m^d$ approximating $W\in\calKC([0;1]^d)$
up to error $2^{-m}$ in Hausdorff distance,
the surface area of the convex hull of $C_m$ approximates 
that of $W$ up to $4d(d-1)\cdot 2^{-m}$.
\IEEEQED\end{enumerate}\end{IEEEproof}

\begin{figure}[htb]
\begin{center}
\includegraphics[width=0.77\columnwidth]{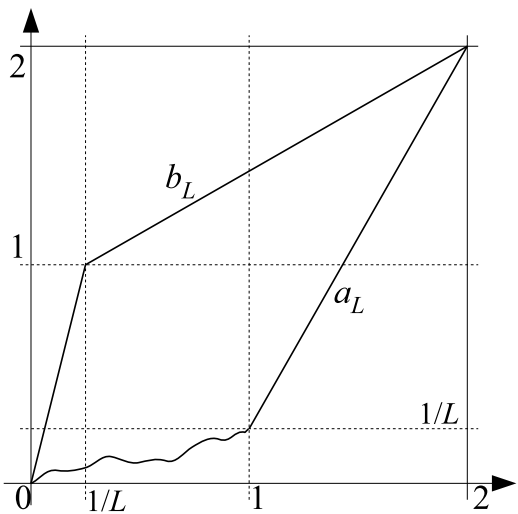}%
\caption{\label{f:Alexander}%
Illustrating the proof of Remark~\ref{r:Frechet}g)}
\end{center}
\end{figure}

\end{document}